\def\SPMWArxiv{1}
\newcommand{\SPMWArxivAuthors}{%
  \author{Hongzheng Chen}
  \affiliation{\institution{Cornell University}\country{USA}}
  \author{Shihan Fang}
  \affiliation{\institution{Cornell University}\country{USA}}
  \author{Niansong Zhang}
  \affiliation{\institution{Cornell University}\country{USA}}
  \author{Sunwoo Kim}
  \affiliation{\institution{Cornell University}\country{USA}}
  \author{Weiran Liu}
  \authornote{Work was done when Weiran interned at Cornell.}
  \affiliation{\institution{Yale University}\country{USA}}
  \author{Zhiru Zhang}
  \affiliation{\institution{Cornell University}\country{USA}}
  \renewcommand{\shortauthors}{Hongzheng Chen et al.}
  \authorsaddresses{}
}

\ifdefined\SPMWArxiv
  \documentclass[acmsmall,screen,nonacm]{acmart}
\else
  \documentclass[acmsmall,screen,review,anonymous,nonacm]{acmart}
\fi

\makeatletter
\@ACM@balancefalse
\makeatother

\usepackage[normalem]{ulem}
\usepackage{color}
\usepackage[]{hyperref}
\usepackage{amsmath,amsfonts}
\usepackage{mathtools}
\usepackage{comment}
\usepackage{booktabs}
\usepackage{multirow}
\usepackage{graphicx}
\usepackage{makecell}
\usepackage{subcaption}
\usepackage{wrapfig}
\usepackage{xcolor}
\usepackage{adjustbox}
\usepackage{siunitx}
\usepackage{xspace}
\usepackage{pifont}

\usepackage[frozencache,cachedir=minted-cache]{minted}
\setminted{fontsize=\scriptsize,autogobble,linenos,xleftmargin=1.5em,numbersep=4pt}
\definecolor{limitation1}{HTML}{FCE5D1}
\definecolor{limitation2}{HTML}{DCEBF7}
\definecolor{limitation3}{HTML}{E0EFDA}
\newcommand{\limitationlabel}[1]{\begingroup\setlength{\fboxsep}{1.4pt}%
  \colorbox{limitation#1}{\strut\textbf{Limitation #1}}\endgroup}
\newcommand{\limitationref}[1]{\begingroup\setlength{\fboxsep}{1.2pt}%
  \colorbox{limitation#1}{\strut\textbf{Limitation~#1}}\endgroup}
\newcommand{\comparisonlinecolor}[2]{%
  \expandafter\def\csname comparisoncolor#1\endcsname{limitation#2}}
\newcommand{\comparisonformatline}[1]{%
  \ifcsname comparisoncolor\number\value{FancyVerbLine}\endcsname
    \edef\FancyVerbHighlightColor{\csname comparisoncolor\number\value{FancyVerbLine}\endcsname}%
  \fi#1}
\AtBeginDocument{%
  \begingroup\fontfamily{zi4}\fontencoding{T1}\selectfont
    \fontencoding{TS1}\selectfont\endgroup
  \DeclareFontShape{T1}{zi4}{m}{it}{<->ssub*zi4/m/n}{}%
  \DeclareFontShape{TS1}{zi4}{m}{it}{<->ssub*zi4/m/n}{}}
\usepackage{tikz}
\usetikzlibrary{fit,backgrounds,calc,positioning,arrows.meta,decorations.pathreplacing}

\AtBeginDocument{%
  \widowpenalty=10000 \clubpenalty=10000 \displaywidowpenalty=10000
  \brokenpenalty=10000 \finalhyphendemerits=1000000}

\newif\ifauthorreview
\ifdefined\SPMWAuthorReview\authorreviewtrue\else\authorreviewfalse\fi
\newcommand{\reviewissue}[2]{\ifauthorreview
  \par\smallskip{\footnotesize\color{red!65!black}\noindent
  \textbf{Author review #1.} #2\par}\smallskip\fi}

\newcommand{\Name}{SPMW\xspace}
\DeclareRobustCommand{\figref}[2][]{Fig.~\hyperref[#2]{\ref*{#2}%
  \if\relax\detokenize{#1}\relax\else(#1)\fi}}

\newcommand{\tbd}[1]{{\color{red!65!black}#1}}
\newcommand{\draftnote}[1]{\ifauthorreview\tbd{#1}\fi}

\AtBeginDocument{%
  }

\ifdefined\SPMWArxiv
  \SPMWArxivAuthors
\fi

\begin{document}
\title[\Name: A Grid-Based Programming Model for Scalable Accelerator Design]{\Name: A Grid-Based Programming Model\texorpdfstring{\\}{ }for Scalable Accelerator Design}

\begin{abstract}
Modern accelerators specialize recurring computation and replicate it across regular hardware structures to achieve high performance. Sustaining this parallelism requires equally deliberate data delivery. Yet design tools often express these structures as sequential programs or flatten them into separate hardware modules and connections. This obscures shared patterns and leads to repeated compiler work.
We introduce Single Program, Multiple Connected Work Units (\Name), a grid-based programming model for scalable accelerator design that makes grid regularity a first-class abstraction.
\Name extends the single program, multiple data (SPMD) abstraction with explicit, typed connection rules, decoupling local work-unit behavior from grid connectivity and data delivery. To our knowledge, \Name is the first programming model to retain regularity throughout the compilation process, combining behavioral HLS with structural RTL generation for both fixed-function and domain-specific instruction-programmable accelerators. We implement \Name with a Python frontend and an MLIR-based compiler that synthesizes each distinct kernel configuration once for reuse across independent hardware instances. Preserving regularity also enables parallel compilation and functional simulation to accelerate iterative design feedback.

We use \Name to generate a $32\times32$ systolic array that reduces GEMM latency and parallel kernel-to-RTL compilation time by $4.36\times$ and $619\times$ relative to AutoSA. Our FFT achieves up to $2.65\times$ lower latency than HP-FFT, a state-of-the-art HLS baseline. Our mini-TPU and our \Name implementation of the FEATHER accelerator demonstrate complex composition and outperform their RTL baselines. Parallel functional simulation provides up to $18.4\times$ speedup over Vitis HLS C simulation on the mini-TPU workloads. \Name is also agent-friendly: given the same design specification, all five coding agents produce \Name designs that satisfy the correctness and latency requirements, with $7.51\times$ lower token usage than for SystemVerilog.
\end{abstract}

\maketitle 

\section{Introduction}
\label{sec:intro}

Modern accelerators rely on structural regularity in computation and communication to achieve high performance within limited area and power budgets. They specialize recurring computations within an application or domain and spatially replicate the resulting processing elements (PEs) for high throughput. For example, 
TPUs replicate multiply-accumulate units in systolic arrays~\cite{jouppi2017tpuv1,jouppi2023tpuv4,jouppi2026tpuv8}. Cerebras's Wafer-Scale Engine connects replicated programmable PEs through a two-dimensional mesh with configurable routes~\cite{lie2023cerebras}. Replication scales parallel execution, while regular communication paths distribute operands and carry results between PEs.

\textbf{Regularity in computation and communication is fundamental to accelerator design.}
Yet, preserving this regularity throughout the design flow remains challenging. Hardware description languages (HDLs) describe circuits at register-transfer level (RTL). A \texttt{generate} loop compactly specifies repeated modules, but compilation expands it into instances and wires. Module hierarchy may survive synthesis without retaining the parameterized grid and connection rule that generated those instances. Full flattening removes the hierarchy itself. High-level synthesis (HLS) raises the abstraction to sequential loop-based programs with scheduling directives~\cite{vitis_hls_2023,google2026xls} and is mostly used to generate application-specific accelerators~\cite{cong2011fpgahls,chen2024understanding,zhang2021fracbnn}. These function-level abstractions leave architectural regularity to the designer. In both cases, compact source can become a bloated intermediate representation (IR) enumerating replicated components and connections. Recovering shared PE behavior, repeated subgraphs, or placement patterns after architectural changes requires additional analysis, delaying validation and design exploration.

\begin{figure}[!t]
\centering
\setminted{fontsize=\fontsize{6.75pt}{8pt}\selectfont}
{\footnotesize
\limitationlabel{1} Replication\qquad
\limitationlabel{2} Topology\qquad
\limitationlabel{3} Data delivery\par}
\vspace{4pt}
\begin{minipage}[t]{0.49\linewidth}
\begin{subfigure}[t]{\linewidth}
\renewcommand{\FancyVerbFormatLine}{\comparisonformatline}
\comparisonlinecolor{3}{2}\comparisonlinecolor{4}{2}
\comparisonlinecolor{25}{2}\comparisonlinecolor{26}{2}
\comparisonlinecolor{16}{1}\comparisonlinecolor{17}{1}
\comparisonlinecolor{18}{1}\comparisonlinecolor{19}{1}
\comparisonlinecolor{21}{1}\comparisonlinecolor{22}{1}
\comparisonlinecolor{23}{1}\comparisonlinecolor{24}{1}
\comparisonlinecolor{27}{1}\comparisonlinecolor{28}{1}
\comparisonlinecolor{20}{3}\comparisonlinecolor{29}{3}
\begin{minted}[highlightlines={3,4,16,17,18,19,20,21,22,23,24,25,26,27,28,29}]{cpp}
using S8 = hls::stream<int8_t>;
using S32 = hls::stream<int32_t>;
void pe(S8 &a_in, S8 &a_out,
        S8 &b_in, S8 &b_out, S32 &c_out) {
#pragma HLS INLINE off
  int32_t acc = 0;
  for (int k = 0; k < K; ++k) {
#pragma HLS PIPELINE II=1
    int8_t a = a_in.read(), b = b_in.read();
    acc += a * b; a_out.write(a); b_out.write(b);
  }
  c_out.write(acc); }
void gemm(int8_t A[M][K], int8_t B[K][N],
          int32_t C[M][N]) {
#pragma HLS DATAFLOW
  S8 a[M][N+1], b[M+1][N]; S32 c[M][N];
#pragma HLS ARRAY_PARTITION variable=a complete dim=0
#pragma HLS ARRAY_PARTITION variable=b complete dim=0
#pragma HLS ARRAY_PARTITION variable=c complete dim=0
  feed_west(A, a); feed_north(B, b);
  for (int i = 0; i < M; ++i) {
#pragma HLS UNROLL
    for (int j = 0; j < N; ++j) {
#pragma HLS UNROLL
      pe(a[i][j], a[i][j+1],
         b[i][j], b[i+1][j], c[i][j]);
    }
  }
  sink_east(a); sink(b); drain(c, C);
}
\end{minted}
\vspace{-6pt}
\caption{C++ HLS with indexed streams~\cite{vitis_hls_2023}.}
\label{subfig:hls}
\end{subfigure}
\end{minipage}
\hfill
\begin{minipage}[t]{0.49\linewidth}
\begin{subfigure}[t]{\linewidth}
\renewcommand{\FancyVerbFormatLine}{\comparisonformatline}
\comparisonlinecolor{3}{2}\comparisonlinecolor{4}{2}
\comparisonlinecolor{5}{2}\comparisonlinecolor{6}{2}
\comparisonlinecolor{10}{2}\comparisonlinecolor{11}{2}\comparisonlinecolor{12}{2}
\comparisonlinecolor{8}{2}\comparisonlinecolor{9}{2}\comparisonlinecolor{26}{1}
\comparisonlinecolor{27}{3}\comparisonlinecolor{28}{3}
\comparisonlinecolor{29}{3}\comparisonlinecolor{30}{3}
\renewcommand{\FancyVerbFormatLine}[1]{%
  \ifnum\value{FancyVerbLine}=14\relax
    \begin{tikzpicture}[overlay]
      \draw[black!65,line width=0.5pt,rounded corners=1pt]
        (-2pt,5.8pt) rectangle
        (\dimexpr\linewidth-1.5em\relax,\dimexpr-7\baselineskip-2pt\relax);
      \node[anchor=north east,font=\sffamily\bfseries\fontsize{6.5pt}{7.4pt}\selectfont,
        inner sep=1pt] at (\dimexpr\linewidth-1.5em-3pt\relax,4.8pt) {Single program};
    \end{tikzpicture}%
  \fi
  \comparisonformatline{#1}}
\begin{minted}[highlightlines={3,4,5,6,8,9,10,11,12,26,27,28,29,30}]{python}
import spmw

class MacIO(spmw.Interface):   # the contract
 west  = spmw.In (int8); north = spmw.In (int8)
 east  = spmw.Out(int8); south = spmw.Out(int8)
 c     = spmw.MemOut(int32) # per-PE result

def mesh(iface, shape):        # the topology
 return spmw.Topology(iface, grid=shape,
  link=lambda i, j: {
   iface.east:  spmw.to((i, j+1), iface.west),
   iface.south: spmw.to((i+1, j), iface.north)})

@spmw.unit
def pe(io: MacIO):
 acc: int32 = 0
 for k in range(K):
  a = io.west.get(); b = io.north.get()
  acc += int(a) * int(b)
  io.east.put(a); io.south.put(b)
 io.c = acc

@spmw.fabric
def gemm(A: int8[M,K], B: int8[K,N],
         C: int32[M,N]):
 P = spmw.place(pe, on=mesh(MacIO, (M, N)))
 t = spmw.steps(K)             # K values per port
 spmw.stream_in(A, into=P.west,  index=(P.rows, t))
 spmw.stream_in(B, into=P.north, index=(t, P.cols))
 spmw.gather   (C, from_=P.c)
\end{minted}
\vspace{-6pt}
\caption{\Name design fragment.}
\label{subfig:spmw}
\end{subfigure}
\end{minipage}
\vspace{-6pt}
\caption{Simplified systolic GEMM construction in C++ HLS and \Name. Colors map the three limitations in \S~\ref{sub:motivation} to the HLS constructs and their \Name counterparts. Python reserves \texttt{from}, so the binding uses \texttt{from\_}.}
\label{fig:comparison}
\Description{Two code panels use matching numbered color categories. Blue highlights indexed PE wiring in HLS and named interfaces and connections in SPMW. Orange highlights HLS stream dimensions and unrolled loops versus the SPMW grid and instance group. Green highlights HLS loaders and sinks versus SPMW boundary bindings. A box labeled Single program encloses SPMW lines 14 through 21, from the unit decorator through the PE body.}
\vspace{-10pt}
\end{figure}


\textbf{Scalable accelerator design requires exploiting regularity throughout the toolflow.} Compiler tools must exploit repeated structures to avoid redundant frontend and backend work, but current toolflows face obstacles at both stages.

\textbf{Challenge 1: Replication and connectivity lack a shared architectural specification.}
Replication alone does not describe a regular accelerator architecture. Single Program, Multiple Data (SPMD) models such as CUDA and Triton organize instances of one program over a logical grid~\cite{cuda,tillet2019triton}. Connectivity and memory supply are expressed separately through communication and memory-access mechanisms, leaving the overall architecture implicit.

This separation also appears in accelerator frontends. The C++ HLS design connects PE instances through indexed streams while loaders separately encode array boundaries and memory accesses, as in \figref[\subref*{subfig:hls}]{fig:comparison}. Resizing the array or partitioning it into independently supplied groups requires coordinated changes across these specifications, even when local arithmetic is unchanged. TAPA~\cite{chi2021tapa,guo2023tapa} replicates tasks through parameterized invocations and connects them through typed streams, but lacks a first-class representation of grid regularity and connection rules. Changes to the interconnection require coordinated edits to stream wiring and memory-access tasks. Even with the help of coding agents, connection indices and loader accesses must be traced across these scattered specifications~\cite{zhang2026pragmas,ding2026abstraction}. This indirection increases an agent's context and editing effort, potentially increasing token use and introducing mismatched connections or memory accesses.


\textbf{Challenge 2: Flattening regular PE arrays hides compilation reuse opportunities.}
On the backend side, tools may repeatedly schedule and synthesize equivalent behavior when they process a regular array as independent kernels. Flattened graphs also leave placement and simulation tools to recover relationships among instances. RapidStream~\cite{guo2022rapidstream} partitions and pipelines dataflow designs for parallel placement and routing. This exploits parallelism across partitions, but does not retain a shared PE program and its replication rule to identify equivalent instances for reuse. Boundaries, constants, and timing contexts can make instances differ, so reuse requires both the original regularity and explicit specialization conditions. Although agents can operate compiler tools~\cite{yang2024sweagent,chen2026magellan,ye2026cake}, long simulation and synthesis runs delay feedback. Generated hardware still needs functional validation~\cite{pinckney2024verilogeval,yu2025spec2rtl}. An error discovered late triggers another edit-and-build cycle, delaying design correction and validation.


\textbf{To address these challenges, we propose \emph{Single Program, Multiple Connected Work Units} (\Name), a programming model that makes regularity a first-class abstraction for scalable accelerator design.} Its central construct is a \emph{grid of connected work units}: a shared component definition is replicated over a logical index space to form processing elements (PEs), and typed connection rules specify how these PEs exchange data. A PE can be a simple processing block or a hierarchy of lower-level components. The grid and connection rules are explicit program objects, available for composition and compiler analysis. Memory bindings connect the resulting boundaries to storage and transfer sequences, making data delivery part of the same description. \Name extends the SPMD principle to explicitly connected PEs, providing a shared architectural specification of replication, connectivity, and data delivery.

The model decouples three design concerns. Work-unit programs define behavior through named, typed ports. Grids and topology define replication and connectivity. Surrounding fabrics define memory organization, boundary supply, and composition. Changing topology can reuse the work-unit program, while changing computation can preserve a compatible interface. A fabric may compose grids with different programs, and units may retain state or interpret instructions. ``Single program'' applies to each replicated family, which allows heterogeneous modules. Grids are logical index spaces and can express nonlocal links and hierarchy.

We implement \Name with a Python frontend and an MLIR flow combining HLS for work-unit behavior and instruction control with structural RTL for the fabric. \textbf{To our knowledge, \Name is the first programming model to preserve grid regularity across the compilation for both fixed-function and domain-specific instruction-programmable accelerators.} Its language-independent architectural contract guides boundary derivation and structural checks before specializing kernels for port usage and coordinate-dependent constants. Each distinct kernel configuration is synthesized once and instantiated as independent hardware modules. The retained structure also enables parallel simulation and physical implementation (\S~\ref{sec:discussion}).

This paper makes the following contributions:
\begin{itemize}
\item We propose \Name, a programming model\ifdefined\SPMWArxiv\footnote{The \Name implementation builds on Allo and is being upstreamed to \url{https://github.com/cornell-zhang/allo}.}\fi{} that makes regularity a first-class abstraction and decouples PE behavior from grid connectivity to construct both fixed-function and domain-specific instruction-programmable accelerators.
\item We implement an MLIR-based compilation flow that retains this regularity to reuse specialized implementations across the grid. It combines behavioral HLS with RTL generation for better structural composition of both accelerator classes, with parallel kernel compilation and functional simulation support to shorten design feedback loops.
\item \looseness=-1 We achieve $4.36\times$ lower GEMM latency and $619\times$ faster parallel kernel-to-RTL compilation than AutoSA with a $32\times32$ systolic array, and up to $2.65\times$ lower FFT latency than HP-FFT. Our mini-TPU and our \Name implementation of FEATHER reduce latency by up to $2.32\times$ and $7.98\times$ versus Gemmini and optimized RTL, respectively. Functional simulation is up to $18.4\times$ faster than Vitis HLS C simulation (csim). All five coding agents' designs pass simulated correctness and latency checks with $7.51\times$ lower median token cost than SystemVerilog.
\end{itemize}

\section{Background and Motivation}
\label{sec:background}
In this section, we review accelerator architectures and design tools, then use a GEMM example to explain the obstacles to architectural changes.

\subsection{Modern Accelerator Architectures}
\label{sub:accelerators}
Accelerators combine repeated execution resources with memory and communication hierarchies. GPU lanes share instruction control within warps, and replicated streaming multiprocessors draw operands from registers, shared memory, and caches~\cite{nvidia2022hopper,cuda}, while general-purpose GPU architectures are beyond this paper's focus on domain-specific accelerators. TPU and AWS Trainium pair systolic matrix engines with on-chip operand storage~\cite{jouppi2017tpuv1,jouppi2023tpuv4,fu2024trainium}. \figref{fig:spatial-architecture} shows TPU-style operand reuse through PE forwarding, with surrounding accumulation and instruction control.

Programmable-core networks provide another organization. Cerebras connects cores and local memories through an on-chip fabric~\cite{lie2023cerebras}. OpenAI's Jalape\~no pairs replicated programmable cores with local memory views and a dedicated collective network~\cite{ho2026jalapeno}. Regular compute and data-delivery structures also appear in Tenstorrent's Blackhole~\cite{tenstorrent}, AMD's AI Engine NPUs~\cite{xdna_npu}, d-Matrix's Corsair~\cite{dmatrix2024corsair}, and HyperAccel's LPU~\cite{moon2024lpu}.

More PEs improve sustained throughput only if operands arrive and results leave fast enough. Memory capacity limits resident tiles, while bank and channel bandwidth constrain concurrent accesses. Hiding off-chip latency requires buffering and scheduled transfers. Grid dimensions, local reuse, and memory organization must therefore be designed together.
\begin{figure}[!t]
\centering
\captionsetup{justification=raggedright,singlelinecheck=false}
\newsavebox{\spmwSyntaxBox}
\newsavebox{\spmwDatapathBox}
\newlength{\spmwFigureHeight}
\sbox{\spmwSyntaxBox}{%
  \begin{minipage}{0.615\linewidth}
\begingroup
\scriptsize
\setlength{\tabcolsep}{0pt}
\renewcommand{\arraystretch}{0.97}
\newcommand{\synkw}[1]{\mathsf{#1}}
\begin{tabular}{@{}l@{}}
Unit $U$\quad Fabric $F$\quad Tensor $A$\quad Instance group $P$ \\
Interface $I$\quad Port $p$\quad Variable $x$\quad Constant $c$
\end{tabular}\par
\noindent
\begin{minipage}[t]{0.49\linewidth}
\begin{tabular}{@{}>{$}l<{$}@{}}
\text{Program }\mathcal{P} ::= \overline{D} \\
\text{Decl. }D ::= \synkw{interface}\ I\ \{\overline{\rho}\} \\
\quad\mid \synkw{unit}\ U(I;\kappa)\ \{\overline{s}\} \\
\quad\mid \synkw{fabric}\ F(I;\kappa)\ \{\overline{a}\} \\
\text{Port }\rho ::= p:\synkw{In}(\tau)\mid p:\synkw{Out}(\tau) \\
\quad\mid p:\synkw{MemIn}(\tau)\mid p:\synkw{MemOut}(\tau) \\
\text{Expr. }e ::= x\mid c\mid e\oplus e\mid e[\overline{e}] \\
\quad\mid p.\synkw{get}()\mid \synkw{helper}(\overline{e}) \\
\text{Unit stmt. }s ::= x=e\mid x[\overline{e}]=e \\
\quad\mid p.\synkw{put}(e)\mid\synkw{if}\ e\ \{\overline{s}\}\ \synkw{else}\ \{\overline{s}\} \\
\quad\mid \synkw{for}\ x\in\synkw{range}(e,e)\ \{\overline{s}\} \\
\quad\mid \synkw{while}\ e\ \{\overline{s}\}
\end{tabular}
\end{minipage}\hfill
\begin{minipage}[t]{0.49\linewidth}
\begin{tabular}{@{}>{$}l<{$}@{}}
\text{Topology }T ::= \synkw{topology}(I,\mathcal{G},\lambda) \\
\text{Component }C ::= U\mid F \\
\text{Source }v ::= A\mid P.p\mid c \\
\text{Fabric stmt. }a ::= P=\synkw{place}(C,T) \\
\quad\mid \synkw{stream\_in}(v,P.p,\phi) \\
\quad\mid \synkw{gather}(A,P.p,\phi) \\
\quad\mid P.p=\synkw{shard}(A,\phi) \\
\quad\mid A=\synkw{mem}(\tau,\kappa)\mid \synkw{copy}(A,A) \\
\quad\mid \synkw{for}\ x\in R\ \{\overline{a}\} \\
\quad\mid \synkw{if}\ k\ \{\overline{a}\}\ \synkw{else}\ \{\overline{a}\}
\end{tabular}
\end{minipage}
\par
\endgroup

  \end{minipage}}
\sbox{\spmwDatapathBox}{\scalebox{0.94}{
\begin{tikzpicture}[
  x=1pt,y=1pt,font=\scriptsize,
  mem/.style={draw=teal!65!black,fill=teal!8,align=center,
              inner sep=3pt,minimum height=19pt},
  pe/.style={draw=blue!60!black,fill=blue!7,minimum width=17pt,
             minimum height=12pt,inner sep=0pt},
  flow/.style={-{Stealth[length=3pt]},semithick},
  lab/.style={font=\scriptsize,align=center}
]
\node[mem,text width=45pt] (dram) at (28,18) {Off-chip memory};
\node[mem,text width=45pt] (bank) at (28,-36) {On-chip buffers / banks};
\node[mem,text width=66pt] (operand) at (112,18) {Operand staging};
\draw[flow] (dram.south) -- (bank.north);
\draw[flow] (dram.east) -- (operand.west);
\coordinate (loaderbus) at (68,-36);
\foreach \r in {0,1,2} {
  \foreach \c in {0,1,2} {
    \node[pe] (p\r\c) at (88+24*\c,-18-18*\r) {PE};
  }
}
\foreach \r in {0,1,2} {
  \foreach \c in {0,1} {
    \pgfmathtruncatemacro{\next}{\c+1}
    \draw[flow,blue!65!black] (p\r\c.east) -- (p\r\next.west);
  }
  \draw[flow,teal!65!black] (loaderbus |- p\r0.west) -- (p\r0.west);
}
\draw[teal!65!black,semithick] (bank.east) -- (loaderbus);
\draw[teal!65!black,semithick] (loaderbus |- p00.west) -- (loaderbus |- p20.west);
\node[lab] at (68,-66) {Loaders};
\foreach \c in {0,1,2} {
  \draw[flow,teal!65!black] (operand.south -| p0\c.north) -- (p0\c.north);
  \foreach \r in {0,1} {
    \pgfmathtruncatemacro{\next}{\r+1}
    \draw[flow,blue!65!black] (p\r\c.south) -- (p\next\c.north);
  }
}
\node[mem,text width=74pt,minimum height=16pt] (result) at (112,-84)
  {Accumulate / collect};
\foreach \c in {0,1,2} {
  \draw[flow] (p2\c.south) -- (result.north -| p2\c.south);
}
\draw[flow] (result.west) -| (bank.south);
\end{tikzpicture}}}
\setlength{\spmwFigureHeight}{\dimexpr\ht\spmwDatapathBox+\dp\spmwDatapathBox\relax}
\ifdim\dimexpr\ht\spmwSyntaxBox+\dp\spmwSyntaxBox\relax>\spmwFigureHeight
  \PackageWarning{spmw-figures}{Abstract syntax exceeds the compact schematic height}
\fi
\typeout{SPMW figure dimensions: schematic \the\wd\spmwDatapathBox\space x \the\spmwFigureHeight; syntax \the\wd\spmwSyntaxBox\space x \the\dimexpr\ht\spmwSyntaxBox+\dp\spmwSyntaxBox\relax}
\begin{minipage}[t]{0.365\linewidth}
\begin{minipage}[t][\spmwFigureHeight][c]{\linewidth}
\centering\usebox{\spmwDatapathBox}
\end{minipage}
\caption{Systolic datapath. \mbox{Control is omitted.}}
\label{fig:spatial-architecture}
\Description{Off-chip memory supplies on-chip buffers and operand staging. Loaders feed a grid of repeated PEs with local forwarding. Collected results return to the buffers.}
\end{minipage}\hfill
\begin{minipage}[t]{0.615\linewidth}
\begin{minipage}[t][\spmwFigureHeight][c]{\linewidth}
\centering\usebox{\spmwSyntaxBox}
\end{minipage}
\caption{Core \Name abstract syntax. Overbars denote sequences, $\oplus$ denotes binary operators, and $R$ is a compile-time iteration domain.}
\label{fig:abstract-syntax}
\Description{Core SPMW abstract syntax begins by identifying names for units, fabrics, tensors, instance groups, interfaces, ports, variables, and constants. Productions define declarations, typed ports, expressions including stream get and helper calls, unit statements including stream put, topologies, components, sources, instance groups, bindings, memories, and structural control flow.}
\end{minipage}
\end{figure}

\subsection{Hardware Design Flow and Programming Models}
\label{sub:progmodels}
Accelerator designers specify execution resources, replication, connections, storage, and control. HDLs such as Verilog describe registers, combinational logic, and their connections at RTL. Chisel~\cite{bachrach2012chisel} generates Verilog from reusable, parameterized hardware components written in Scala.

HLS compiles behavior and scheduling directives to RTL, then logic synthesis and physical implementation map it to FPGA resources~\cite{vitis_hls_2023} or ASIC standard cells~\cite{siemens2026catapult}. Vitis HLS streams provide typed FIFO channels for passing values between functions~\cite{vitis_hls_2023}. SYCL's oneAPI FPGA implementation connects kernels through pipes~\cite{khronos_sycl2020,intel_oneapi_pipes}. TAPA~\cite{guo2023tapa} expresses FPGA accelerators as hierarchies of concurrent tasks connected through typed streams. Spatial~\cite{koeplinger2018spatial} expresses nested parallel loops and reductions and automatically banks on-chip memories to serve concurrent accesses. Allo~\cite{chen2024allo} decouples algorithms from hardware customization and composes customized kernels through schedules. These systems organize design around modules, tasks, or computation transformations. We discuss additional DSL and compiler approaches in \S~\ref{sec:related}.

\reviewissue{R12}{The novelty claim is the compiler-visible contract preserving the relationship among a shared work unit, parameterized connections, and boundary delivery for structural checking, specialization, and reuse. RTL and hardware generators already support parameterized arrays and instance specialization, and TAPA/HIDA expose task connectivity and hierarchy. The introduction explains the preserved relationship and its use during compilation. Detailed named-system comparisons remain in Related Work. This is not a claim of exclusive circuit expressivity or complete functional verification.}

\subsection{A Motivating Example}
\label{sub:motivation}
\figref{fig:comparison} compares output-stationary GEMM descriptions in C++ HLS and our Python realization of \Name. In the shown array, each PE column computes a different column of \texttt{C}. Values from each row of \texttt{A} enter at the west boundary and pass through all PEs in that row, where they are multiplied by different columns of \texttt{B}. Suppose the designer divides the PE columns into two groups to shorten the horizontal forwarding paths. The PE program and PE count stay unchanged, but each row gains a new input for the second group and an output to drain from the first. Both groups must receive the same \texttt{A} values. This split requires coordinated edits to channel storage for the replicated PEs, indexed PE connections, and boundary loaders and sinks. We summarize three limitations.

\textbf{\limitationlabel{1}: replication spans multiple constructs.}
Array shapes and partition directives (L16--L19) define storage for the channels, while unrolled loops replicate PE calls (L21--L28). Turning off PE inlining (L5) does not prevent repeated scheduling. Shared constants parameterize the dimensions, but the extra boundary channels must still match the PE instances created by these loops. The grid is implicit across declarations and control flow. Flattening can obscure the common PE program and grid, leaving downstream compilation to rediscover opportunities for reuse.

\textbf{\limitationlabel{2}: topology is implicit in indexed wiring.}
The PE has reusable stream ports (L3--L4), while the PE call encodes neighbors through array subscripts and argument order (L25--L26). At the split, the second group's first PE must read the new boundary stream. A stale index can leave a PE attached to its former neighbor or without an input source, causing incorrect results or a stalled stream.

\textbf{\limitationlabel{3}: rewiring PEs leaves operand delivery unchanged.}
Loaders feed the west and north boundaries (L20), while sinks and a drain consume outputs (L29). The loader must supply the same \texttt{A} values to both groups with sufficient bandwidth, and a sink must drain the first group's new output. Editing PE calls alone does not update these boundary operations.

\Name connects these design choices through the declarations in \figref[\subref*{subfig:spmw}]{fig:comparison}. The designer instantiates a shared PE program (L14--L21) on a grid whose shape parameterizes the topology (L26). The resulting \emph{instance group} \texttt{P} names these logical instances and their ports. Physical placement assigns their chip locations later. The compiler derives the PE instances and channels together, retaining their shared structure for compilation and addressing \limitationref{1}. The PE uses named ports from the interface contract \texttt{MacIO} (L3--L6), while a separate connection rule specifies their neighbors (L8--L12). This makes topology explicit and allows it to change without rewriting the PE program (\limitationref{2}). The instance group also exposes the boundary bundles used by input bindings and result collection (L27--L30), linking data delivery to the declared topology (\limitationref{3}). To form the two column groups, the designer only needs to remove the horizontal links between them and bind both groups' west inputs to the same rows of \texttt{A}, while retaining the PE program.

C++ generators in Vitis HLS~\cite{vitis_hls_2023}, Catapult HLS~\cite{siemens2026catapult}, and SYCL~\cite{khronos_sycl2020} can automate these edits, but must specify how topology changes update channels, PE connections, and loader accesses. For example, splitting a chain creates a new input that must be supplied. \Name retains grids, connections, and boundary bindings so the compiler can derive instances and exposed ports for structural checks and kernel reuse. \S~\ref{sub:attnpv} develops a detailed splitting example for attention.

\section{The \Name Programming Model}
\label{sec:model}
\Name decouples \emph{what} each work unit executes from \emph{how} units are assembled and supplied with data. A \emph{unit} defines local computation. A \emph{fabric} defines its hardware organization. \emph{Elaboration} evaluates this structure at compile time to construct instances, connections, memories, and bindings. Static parameters, such as the GEMM dimensions \texttt{M}, \texttt{N}, and \texttt{K} in \figref[\subref*{subfig:spmw}]{fig:comparison}, are fixed per build but may change between builds. Array contents and instruction streams arrive at run time. \figref{fig:abstract-syntax} shows the abstract syntax of \Name and distinguishes run-time unit statements $s$ from elaboration-time fabric statements $a$.

\begin{figure}[!t]
\centering
\begin{tikzpicture}[
  x=0.78cm, y=0.90cm,
  font=\fontsize{7.5pt}{8.5pt}\selectfont,
  head/.style={font=\fontsize{8pt}{9pt}\selectfont\bfseries,
               anchor=west,align=left,text width=3.6cm},
  unit/.style={draw=black!65, fill=orange!12, minimum width=0.55cm,
               minimum height=0.48cm, inner sep=1pt},
  socket/.style={rectangle, draw=blue!70!black, fill=white,
                 minimum size=4pt, inner sep=0pt},
  stream/.style={-{Stealth[length=3.5pt]}, thick, blue!70!black},
  vertical/.style={stream, red!65!black},
  binding/.style={circle, draw=blue!70!black, fill=white,
                  minimum size=4pt, inner sep=0pt},
  rule/.style={fill=black!4, rounded corners=1pt, inner sep=4pt},
  frame/.style={draw=blue!45!black, rounded corners=2pt, fill=blue!2},
  mem/.style={draw=green!45!black, fill=green!12, minimum size=4pt, inner sep=0pt}
]
\node[head] at (0,0) {(a) Unit: behavior\\and interface};
\node[head,text width=4.2cm] at (5.35,0) {(b) Fabric: typed\\instance groups and links};
\node[head] at (11.75,0) {(c) Fabric as a component};

\begin{scope}[yshift=-6pt]
\node[unit, minimum width=1.45cm, minimum height=0.90cm] (u)
  at (2.05,-1.38) {$U=\texttt{pe}$};
\foreach \side in {west,east,north,south} {
  \node[socket] at (u.\side) {};
}
\draw[stream] (0.68,-1.38) -- (u.west);
\draw[stream] (u.east) -- (3.40,-1.38);
\draw[vertical] (2.05,-0.62) -- (u.north);
\draw[vertical] (u.south) -- (2.05,-2.15);
\node[above, blue!70!black] at (0.72,-1.38) {\texttt{west}};
\node[above, blue!70!black] at (3.27,-1.38) {\texttt{east}};
\node[above, red!65!black] at (2.05,-0.62) {\texttt{north}};
\node[below, red!65!black] at (2.05,-2.15) {\texttt{south}};
\node[mem] (result) at (2.78,-1.68) {};
\draw[-{Stealth[length=3pt]}, green!45!black] (result) -- (3.50,-1.92);
\node[right, green!35!black] at (3.48,-1.92) {\texttt{c}};
\node at (2.1,-2.75) {$I=\texttt{MacIO}$};
\node[align=center, text width=3.4cm] at (2.1,-3.25)
 {Streams: \texttt{int8}\\\texttt{c}: \texttt{MemOut(int32)}};
\end{scope}

\draw[frame] (5.45,-0.45) rectangle (11.25,-4.20);
\node at (8.35,-0.78) {$F:\quad T=(I,\mathcal{G},\lambda),\ I=\texttt{MacIO}$};
\foreach \r in {0,1} {
 \foreach \c in {0,1} {
  \node[unit] (p\r\c) at (7.10+2.00*\c,-1.52-0.87*\r) {$U$};
  \node[socket] at (p\r\c.west) {};
  \node[socket] at (p\r\c.east) {};
 }
 \draw[stream] (p\r0.east) -- (p\r1.west);
 \node[binding] (feed\r) at (6.40,-1.52-0.87*\r) {};
 \draw[stream] (feed\r) -- (p\r0.west);
}
\foreach \c in {0,1} {
 \draw[vertical] (p0\c.south) -- (p1\c.north);
}
\draw[-{Stealth[length=4pt]}, dashed, black!60]
 (3.90,-1.52) -- (6.00,-1.52);
\node[above, align=center] at (4.95,-1.45) {same $I$};
\node[rule] at (8.35,-3.03) {\texttt{P = spmw.place(U, on=T)}};
\node[draw=red!65!black, fill=red!4, rounded corners=1pt,
      inner sep=3pt] (bad) at (6.84,-3.72) {$U':\texttt{OtherIO}$};
\draw[red!65!black, thick] (bad.east) -- (8.05,-3.72);
\draw[red!65!black, very thick] (7.93,-3.59) -- (8.17,-3.85)
                                    (7.93,-3.85) -- (8.17,-3.59);
\node[right, red!65!black] at (8.28,-3.72) {mismatch: rejected};

\draw[frame] (11.95,-0.45) rectangle (17.10,-4.20);
\node at (14.52,-0.78) {\texttt{tiled\_gemm}: tile $(i,j)$};
\draw[draw=green!40!black, fill=white, rounded corners=2pt]
 (12.55,-1.10) rectangle (16.45,-3.18);
\node at (14.52,-1.31) {\texttt{tile\_gemm}};
\node[mem] (a) at (12.55,-1.93) {};
\node[mem] (b) at (12.55,-2.70) {};
\node[mem] (c) at (16.45,-2.24) {};
\node[left] at (12.42,-1.93) {$a$};
\node[left] at (12.42,-2.70) {$b$};
\node[right] at (16.57,-2.24) {$c$};
\node[draw=blue!65!black, fill=white, minimum width=0.78cm,
      minimum height=0.46cm, align=center, inner sep=1pt]
 (abank) at (13.33,-1.93) {$A$ banks};
\node[draw=red!65!black, fill=white, minimum width=0.78cm,
      minimum height=0.46cm, align=center, inner sep=1pt]
 (bbank) at (13.33,-2.70) {$B$ banks};
\draw[stream] (a) -- (abank.west);
\draw[vertical] (b) -- (bbank.west);
\coordinate (abus) at ([xshift=3pt]abank.east);
\coordinate (bbus) at ([xshift=6pt]bbank.east);
\coordinate (btop) at (14.70,-1.57);
\foreach \r in {0,1} {
 \foreach \j in {0,1} {
  \node[unit, minimum width=0.40cm, minimum height=0.35cm]
   (q\r\j) at (14.70+0.90*\j,-1.93-0.67*\r) {$U$};
 }
 \draw[stream] (abank.east) -- (abus) |- (q\r0.west);
 \draw[stream] (q\r0.east) -- (q\r1.west);
}
\draw[red!65!black,preaction={draw=white,line width=2pt}]
 (bbank.east) -- (bbus) -- (bbus |- btop) -- (15.60,-1.57);
\foreach \j in {0,1} {
 \draw[vertical] (14.70+0.90*\j,-1.57) -- (q0\j.north);
 \draw[vertical] (q0\j.south) -- (q1\j.north);
}
\draw[green!45!black] (q01.east) -- (16.03,-1.93) -- (16.03,-2.60) -- (q11.east);
\draw[-{Stealth[length=3pt]},green!45!black] (16.03,-2.24) -- (c);
\node[align=center, text width=3.6cm] at (14.52,-3.73)
 {\texttt{TileIO}: $a,b$ inputs, $c$ output\\Internal units use \texttt{MacIO}};

\end{tikzpicture}
\caption{Typed composition of units and fabrics. (a) A unit declares its local behavior and ports. (b) A fabric instantiates it on a topology with a matching interface and rejects an incompatible component. (c) \texttt{tile\_gemm} exposes memory ports to \texttt{tiled\_gemm} while encapsulating its PE grid and storage. \figref{fig:tiled} gives the code.}
\label{fig:topology-design}
\Description{Three panels show a MAC unit with typed stream and memory ports, its compatible instance group within a fabric with rejection of another interface, and the tile_gemm child inside tiled_gemm. The child exposes memory inputs a and b and memory output c. Its internal A row banks and B column banks supply the MacIO PE mesh.}
\vspace{-22pt}
\end{figure}
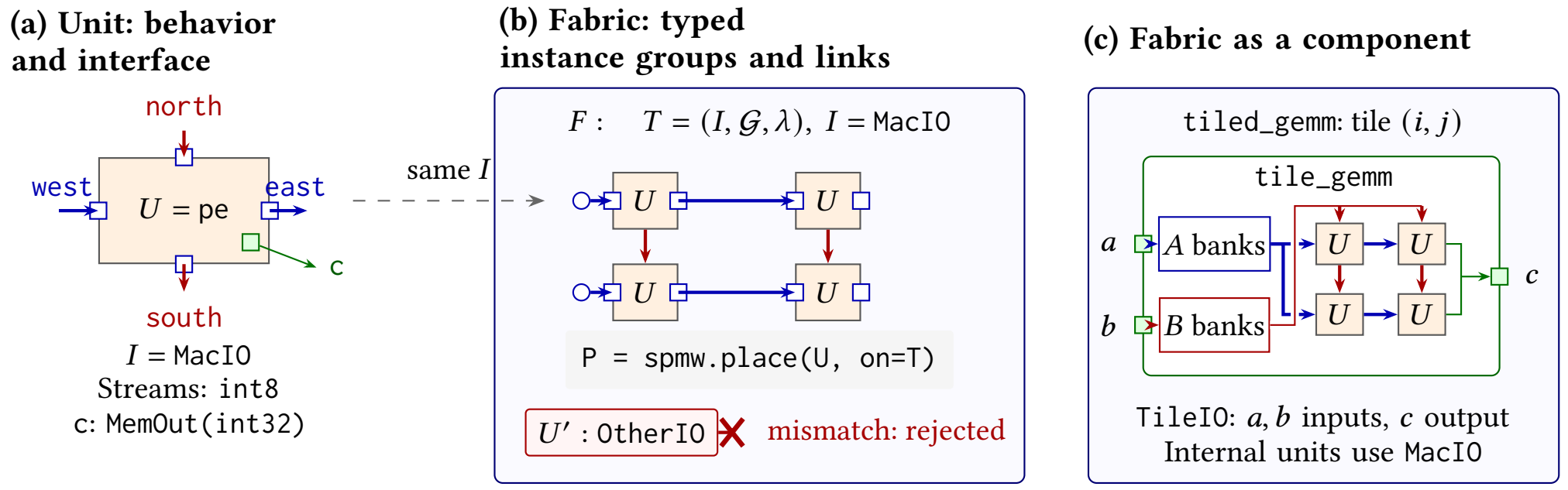

\subsection{Interfaces and Units}
\label{sub:overview}\label{sub:interface}
An \emph{interface} $I$ is a finite set of named ports. Each port declares its protocol, access direction, payload type, and shape. The protocol distinguishes ordered FIFO transfers from indexed memory access, independently of direction. In the frontend, a \texttt{spmw.Interface} class defines this contract. \texttt{spmw.In(T)} and \texttt{spmw.Out(T)} declare stream inputs and outputs with payload type \texttt{T}, which may be a scalar or a statically shaped array. \texttt{spmw.MemIn(T)} and \texttt{spmw.MemOut(T)} declare memory inputs and outputs with the same type notation. \texttt{p.get()} removes and returns the next input value, blocking on an empty FIFO. \texttt{p.put(e)} sends \texttt{e} through the output port, blocking until the FIFO has capacity. Memory ports obey their declared access permissions. Undeclared ports, wrong-direction accesses, and stream operations on memory ports are invalid even when payload types match. Port types specify no transfer rate, latency, or initiation interval. \texttt{depth} specifies FIFO capacity (\S~\ref{sub:topology}).

A \emph{unit} is a leaf component $U=(I,\sigma,B;\kappa)$, with interface $I$, private state $\sigma$, imperative body $B$, and static parameters $\kappa$, such as the GEMM kernel dimension \texttt{K}. Private state may include local memory and registers. The \texttt{@spmw.unit} decorator declares the component, whose interface parameter exposes exactly the ports in $I$, as shown in \figref[a]{fig:topology-design}. In \figref[\subref*{subfig:spmw}]{fig:comparison}, \texttt{pe(io: MacIO)} reads and forwards two streams, accumulates their products, and writes the result (L14--L21).

A unit must be \emph{self-contained}. Run-time inputs and externally visible effects pass through its ports, and other mutable storage belongs to the instance. Local variables, control flow, and helpers obey this rule. Types and immutable elaboration-time constants, such as \texttt{int32} and \texttt{K}, accompany specialization. Capturing fabric arrays or instance-group handles to bypass the interface, accessing another instance's ports, or sharing mutable Python globals violates the contract.

Each instance owns its state, including the GEMM accumulator \texttt{acc}, initialized on entry and updated across \texttt{K} iterations. Instances may respond independently to run-time data or instructions. Unit bodies cannot execute structural operations such as \texttt{spmw.place} (\S~\ref{sub:topology}), declare fabric memories, or change connections at run time.

\begin{figure}[!t]
\begin{minipage}[t]{0.49\linewidth}
\vspace{0pt}
\begin{minted}{python}
M, N, K = 512, 512, 512
Rt, Ct = 16, 16
TM, TN = M // Rt, N // Ct
class TileIO(spmw.Interface):
 a = spmw.MemIn(int8[Rt, K])
 b = spmw.MemIn(int8[K, Ct])
 c = spmw.MemOut(int32[Rt, Ct])
@spmw.fabric
def tile_gemm(io: TileIO):
 P = spmw.place(pe, on=mesh(MacIO, (Rt, Ct)))
 a_stage = spmw.mem(int8[Rt, K],
     layout=spmw.banked(on="row"))
 b_stage = spmw.mem(int8[K, Ct],
     layout=spmw.banked(on="col"))
 spmw.copy(io.a, into=a_stage)
 spmw.copy(io.b, into=b_stage)
 t = spmw.steps(K)
 spmw.stream_in(a_stage, into=P.west,
                index=(P.rows, t))
 spmw.stream_in(b_stage, into=P.north,
                index=(t, P.cols))
 spmw.gather(io.c, from_=P.c)
\end{minted}
\end{minipage}\hfill
\begin{minipage}[t]{0.49\linewidth}
\vspace{0pt}
\begin{minted}[firstnumber=23]{python}
@spmw.fabric
def tiled_gemm(A: int8[M,K],
               B: int8[K,N],
               C: int32[M,N]):
 T = spmw.place(tile_gemm,
     on=spmw.Topology(TileIO, grid=(TM, TN)))
 spmw.shard(A, into=T.a, dim=0)
 spmw.shard(B, into=T.b, dim=1)
 spmw.shard(C, from_=T.c)
\end{minted}
\par\smallskip
\begin{tikzpicture}[
  x=1cm, y=1cm, font=\fontsize{7pt}{8pt}\selectfont,
  view/.style={dashed, -{Stealth[length=2.8pt]}, line width=0.45pt},
  flow/.style={-{Stealth[length=2.8pt]}, line width=0.5pt},
  tile/.style={draw=green!45!black, fill=green!7,
               minimum width=0.65cm, minimum height=0.59cm, inner sep=1pt},
  pe/.style={draw=orange!70!black, fill=orange!15,
             minimum size=0.20cm, inner sep=0pt}
]
\path[use as bounding box] (0,0) rectangle (4.55,-3.45);
\node[align=center] at (1.05,-0.25) {\texttt{tiled\_gemm}\\Tile grid (excerpt)};
\node[red!65!black] at (0.83,-0.82) {$B_0$};
\node[red!65!black] at (1.65,-0.82) {$B_1$};
\foreach \i in {0,1} {
 \node[blue!70!black] at (0.12,-1.45-0.88*\i) {$A_{\i}$};
 \foreach \j in {0,1} {
  \node[tile] (t\i\j) at (0.83+0.82*\j,-1.45-0.88*\i) {$C_{\i\j}$};
 }
 \draw[view,blue!70!black] (0.12,-1.62-0.88*\i)
  -- (0.12,-1.84-0.88*\i) -- (0.64,-1.84-0.88*\i)
  -- (0.64,-1.745-0.88*\i);
 \draw[view,blue!70!black] (0.64,-1.84-0.88*\i)
  -- (1.46,-1.84-0.88*\i) -- (1.46,-1.745-0.88*\i);
}
\foreach \j in {0,1} {
 \draw[view,red!65!black] (0.83+0.82*\j,-1.00)
  -- (0.83+0.82*\j,-1.155);
 \draw[view,red!65!black] (0.83+0.82*\j,-1.05)
  -- (1.21+0.82*\j,-1.05) -- (1.21+0.82*\j,-2.33)
  -- (1.155+0.82*\j,-2.33);
}
\draw[densely dotted, black!45] (2.27,-0.05) -- (2.27,-3.40);
\begin{scope}[shift={(2.45,4.0)}]
\node[align=center] at (1.05,-4.22) {Each \texttt{tile\_gemm}\\\texttt{TileIO}};
\draw[draw=green!45!black, rounded corners=1pt, fill=green!2]
 (0.07,-4.70) rectangle (2.06,-7.03);
\node[blue!70!black] at (0.55,-4.52) {$a$};
\node[red!65!black] at (1.56,-4.52) {$b$};
\node[draw=blue!60!black, fill=white, align=center,
      minimum width=0.80cm, inner sep=1pt,
      font=\fontsize{6.5pt}{7.5pt}\selectfont]
 (abanks) at (0.55,-5.04) {row\\banks};
\node[draw=red!65!black, fill=white, align=center,
      minimum width=0.80cm, inner sep=1pt,
      font=\fontsize{6.5pt}{7.5pt}\selectfont]
 (bbanks) at (1.56,-5.04) {column\\banks};
\draw[flow,blue!70!black] (0.55,-4.68) -- (abanks.north);
\draw[flow,red!65!black] (1.56,-4.68) -- (bbanks.north);
\foreach \r in {0,1} {
 \foreach \j in {0,1} {
  \node[pe] (p\r\j) at (0.72+0.65*\j,-5.87-0.46*\r) {};
 }
 \draw[flow,blue!70!black] (p\r0.east) -- (p\r1.west);
}
\draw[blue!70!black] (abanks.south)
 -- (0.55,-5.52) -- (0.27,-5.52) -- (0.27,-6.33);
\draw[red!65!black] (bbanks.south) -- (1.56,-5.57) -- (0.72,-5.57);
\foreach \j in {0,1} {
 \draw[flow,red!65!black] (p0\j.south) -- (p1\j.north);
 \draw[flow,red!65!black] (0.72+0.65*\j,-5.57) -- (p0\j.north);
}
\foreach \r in {0,1} {
 \draw[flow,blue!70!black] (0.27,-5.87-0.46*\r) -- (p\r0.west);
}
\node at (1.05,-6.68) {PE array};
\draw[flow,green!45!black] (1.05,-6.87) -- (1.05,-7.19);
\node[green!35!black, below] at (1.05,-7.17) {$c$};
\end{scope}
\end{tikzpicture}
\end{minipage}
\vspace{-6pt}
\caption{Hierarchical GEMM for \figref[c]{fig:topology-design}. \texttt{TileIO} exposes child memory ports. Dashed bindings share $A_i$ by row and $B_j$ by column. Each tile banks its operands and owns $C_{ij}$.}
\label{fig:tiled}
\Description{Code defines TileIO with memory inputs a and b and output c, a child tile_gemm with row-banked A staging and column-banked B staging, and its tiled_gemm parent. Beside the code, a two-by-two tile grid shows shared row views of A, shared column views of B, and disjoint output blocks C00, C01, C10, and C11. An expanded tile shows a copied into row banks, b copied into column banks, both feeding the PE array, and results leaving through c.}
\vspace{-4pt}
\end{figure}
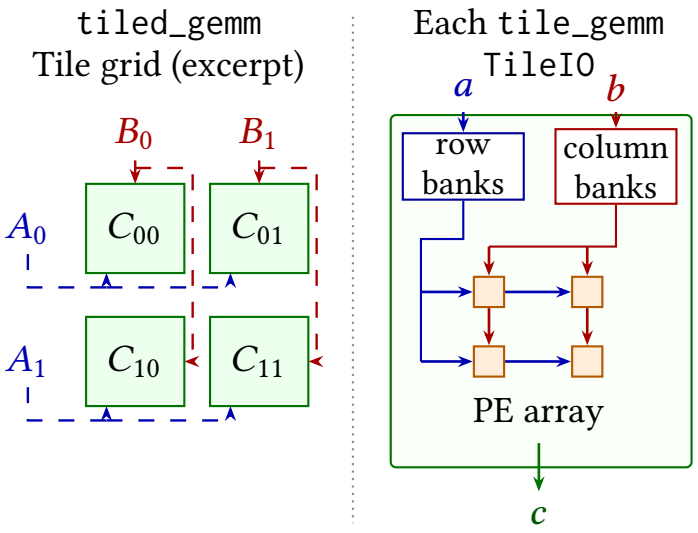

\subsection{Fabrics and Structural Elaboration}
\label{sub:unitfabric}\label{sub:place}\label{sub:hierarchy}
A \texttt{@spmw.fabric} body runs during elaboration to construct hardware from component ports and array references. It uses static arithmetic to calculate shapes, coordinates, and binding maps, but does not execute unit bodies or compute on run-time values.

Given a fabric $F$ and static design parameters $\kappa$ whose values are chosen for this build, elaboration produces a finite hardware graph
\[
  \operatorname{Elab}(F;\kappa)=(\mathcal{V},\mathcal{E},\mathcal{M},\beta),
\]
where $\mathcal{V}$ denotes logical unit instances, $\mathcal{E}$ contains their connections, $\mathcal{M}$ describes memories, and $\beta$ records external and memory bindings. Standalone storage in $\mathcal{M}$, such as the A/B banks in \figref[c]{fig:topology-design}, is declared with \texttt{spmw.mem} (\S~\ref{sub:bindings}) and need not belong to a PE. A memory binding associates a component port with a tensor view specifying which elements it may read or write. The compiler uses this association in $\beta$ to generate access paths and address mappings.

This graph describes hardware semantics. The IR retains each regular family as an instance group referencing one shared unit function, without copying its body at every coordinate. Retained connection rules let the compiler recompute links and exposed ports after architectural changes.

A fabric loop generates repeated structure. A unit loop describes repeated work on the generated hardware, subject to scheduling. Grid sizes, component choices, and connection rules must therefore be determined without run-time stream or array values.

A fabric can expose a typed interface and become a component in a parent fabric. In \figref{fig:tiled}, \texttt{tiled\_gemm} replicates \texttt{tile\_gemm} on a $T_M\times T_N$ grid and binds its \texttt{TileIO} ports to tensor views (L23--L31). Each child contains an $R_t\times C_t$ PE mesh with local buffers and stream bindings (L8--L22). The parent uses only the exposed interface, so the child can change its banking or connections independently. This composition rule applies at further levels.
\subsection{Grids and Typed Connectivity}
\label{sub:topology}
A logical $d$-dimensional grid with $n_a$ coordinates on axis $a=0,\ldots,d-1$ is the Cartesian product
\[
 \mathcal{G}=\prod_{a=0}^{d-1}\{0,\ldots,n_a-1\},\qquad d\geq1,\ n_a>0.
\]
For example, an $M\times N$ mesh has $d=2$, $n_0=M$, and $n_1=N$, with coordinates $(i,j)$.

A topology $T=(I,\mathcal{G},\lambda)$ combines an interface with this grid and a connection rule. For the coordinate form used in the paper, $\lambda$ is a partial function from a source coordinate and stream output to a destination coordinate and stream~input:
$
 \lambda:\mathcal{G}\times\operatorname{Out}(I)
   \rightharpoonup\mathbb{Z}^{d}\times\operatorname{In}(I).
$
In \figref[\subref*{subfig:spmw}]{fig:comparison}, the \texttt{link} function returns these output-to-destination pairs using \texttt{spmw.to} (L11--L12). It depends only on coordinates and static parameters. An in-grid destination creates a channel. An omitted mapping or an off-grid destination leaves the output exposed at the boundary.

Every edge must name ports owned by the topology's interface, connect an output to an input, and preserve payload type and shape. Each destination input has at most one producer. Connecting two outputs to the same input is invalid because the rule supplies no implicit arbitration or reduction. A coordinate link is point-to-point. Broadcast or reduction requires calling an explicit communication operation (\S~\ref{sub:streambindings}). Fabric bindings associate memory ports with tensor views through \texttt{spmw.shard}, or collect memory outputs into a destination tensor through \texttt{spmw.gather} (\S~\ref{sub:delivery}). Coordinate rules create FIFO channels only between stream ports.

FIFO capacity is chosen during elaboration. A stream port may request a capacity with \texttt{depth}. A link uses the largest explicit request at its endpoints, or two values if neither specifies one. Reads block on an empty FIFO and writes block on a full FIFO, propagating backpressure between units. The designer can increase capacity to accommodate pipeline overlap or cyclic dependencies. \Name does not infer a minimum deadlock-free depth.

The operation \texttt{spmw.place(C, on=T)} instantiates the explicitly supplied component $C$ on topology $T$. The component may be a unit or a fabric with an exposed interface. The frontend checks that $C$ declares $T$'s named interface $I$, then creates one independent instance per grid coordinate. Interface matching validates $C$. Distinct interface declarations remain incompatible even with identical port names. \figref[a,b]{fig:topology-design} shows an accepted match and a rejected mismatch. Thus, \texttt{io.east} in the unit and \texttt{iface.east} in the link rule refer to the same port declaration. In \figref[\subref*{subfig:spmw}]{fig:comparison}, the mesh connects east to west and south to north (L8--L12).

Logical grid coordinates identify instances without prescribing nearest-neighbor connectivity or a physical floorplan. Rules can describe meshes, stage permutations, and nonlocal links. Several instance groups may use different unit programs and connect through compatible boundaries, allowing heterogeneous accelerators. Elaboration resolves connectivity before physical placement.

\subsection{Data Delivery}
\label{sub:delivery}\label{sub:boundary}
Data delivery connects exposed ports to operand sources and result destinations. A \emph{boundary bundle} collects the exposed instances of one declared port within an instance group, indexed by grid coordinates. For example, \texttt{P.west} collects the west input at the start of each mesh row. It groups endpoints of one port, rather than whole interfaces or all same-typed ports in a fabric. Both stream and tensor-memory bindings use these bundles.

Formally, let $\mathcal{E}_T$ be the in-grid edges derived from $\lambda$. An endpoint $(x,p)$ pairs a coordinate $x$ with a port $p$, in the same order used by $\lambda$. For an input port $p$, the instance group exposes the bundle
\[
 \operatorname{Bnd}_{p}(T)=\{(x,p)\mid x\in \mathcal{G},\ (x,p)\notin\operatorname{dst}(\mathcal{E}_T)\}.
\]
Here, $\operatorname{dst}(\mathcal{E}_T)$ contains all destination endpoints in $\mathcal{E}_T$, so exposed inputs have no internal producer. Output bundles analogously contain outputs without an internal consumer. Cutting a horizontal link adds its destination's west input to \texttt{P.west}, even if that coordinate lies inside the grid. Memory ports likewise form bundles indexed by grid coordinates and are bound to tensor views.

Every consumed input must have exactly one source, supplied by an internal channel or fabric binding. Residual inputs cannot be left unspecified or silently treated as zero. Unused forwarding outputs may be discarded. The elaboration-time structural checker complements computation type checking by validating port accesses and topology constraints. However, well-formed connectivity alone proves neither deadlock freedom nor numerical correctness. In the following, we introduce two kinds of bindings.

\subsubsection{Stream Bindings}
\label{sub:streambindings}
Stream bindings specify how values enter and leave a grid over time, independently of unit computation. The \texttt{spmw.stream\_in} primitive supplies an input bundle from a memory array, a constant, or a compatible output bundle. For a memory source, an \texttt{index} map selects elements by endpoint coordinates and transfer step. The primitive \texttt{spmw.steps(K)} names $K$ ordered transfers. Generated loaders advance when their output streams can accept values.

For whole-bundle transfers, \Name provides \texttt{spmw.scatter} to distribute tensor elements to instances and \texttt{spmw.gather} to collect their outputs into a tensor. Both support stream and memory ports. A coordinate map assigns each output's destination in \texttt{gather}, while a temporal index maps successive stream values.

In \figref[\subref*{subfig:spmw}]{fig:comparison}, \texttt{t = spmw.steps(K)} defines $t=0,\ldots,K-1$ (L27). The binding \texttt{spmw.stream\_in(A, into=P.west, index=(P.rows,t))} sends \texttt{A[r,t]} to the boundary input in row $r$ (L28). The north binding sends \texttt{B[t,c]} to column $c$ (L29). Finally, \texttt{spmw.gather(C, from\_=P.c)} collects each PE's scalar memory output into the corresponding element of~\texttt{C}~(L30). For a stream source, \texttt{spmw.stream\_in} pairs the source output bundle with the destination input bundle in coordinate order. Their counts, payload types, and shapes must match. This binding preserves value order and requires no memory loader. The same unit can therefore receive operands from memory loaders or other units without changing its arithmetic.

\subsubsection{Tensor-Memory Bindings}
\label{sec:memory}\label{sub:bricks}\label{sub:bindings}
Tensor sharding assigns tensor views to component memory ports. The compiler records \texttt{spmw.shard} mappings during elaboration to generate access paths from storage to component instances. Views must match port types and shapes and stay within tensor bounds. Read-only views may overlap, while output views must be disjoint and cover the result with one writer per element. \Name applies tensor partitioning and replication over logical grids from Alpa~\cite{zheng2022alpa} and other tensor-sharding work~\cite{xu2021gspmd,sami2025partir,shardy,chen2024slapo} to hardware memory ports.

Banking physically partitions a buffer for concurrent access. The \texttt{banked} layout of \texttt{spmw.mem} creates independent banks, while \texttt{spmw.shard} maps components to tensor elements. Both construct hardware at compile time. Explicit copies or generated data movers load values at run time, allowing a banked weight buffer to receive new tiles without changing its structure.
Bank conflicts occur when simultaneous accesses exceed a bank's ports, while sharding supplies no extra memory ports or run-time arbiter between independent components, which require separate banks or replicas. Our structural checker rejects multiple writers or concurrent read/write to one declared memory.

In \figref{fig:tiled}, \texttt{spmw.shard} shares each row block of $A$ within a tile row and each column block of $B$ within a tile column (L29--L30). Each tile owns a distinct output block in $C$ (L31). These bindings connect parent tensors to child memory ports. Each child declares row-banked \texttt{a\_stage} and column-banked \texttt{b\_stage} (L11--L14), populated by explicit copies (L15--L16). The \texttt{stream\_in} bindings supply the west and north boundaries concurrently (L18--L21). Staging can be bypassed if the source provides sufficient bandwidth, but sharing a view does not create physical read ports.

\section{Case Studies}
\label{sec:case}
We use three cases to demonstrate \Name's capabilities. (1)~\textbf{Generality.} We show how \Name supports different systolic-array dataflows by changing PE computation and connection patterns. (2)~\textbf{Programmability.} We then prototype a mini-TPU that executes run-time programs by adding instruction control to the systolic array as its matrix unit. (3)~\textbf{Flexibility.} We further prototype a grouped attention array by changing topology and data delivery while reusing its arithmetic units, as an example of rapid hardware exploration.
\begin{figure}[t]
\begin{minipage}[t]{0.49\linewidth}
\vspace{0pt}
\begin{minted}{python}
class WsIO(spmw.Interface):
 a_in = spmw.In (int8);  a_out = spmw.Out(int8)
 p_in = spmw.In (int32); p_out = spmw.Out(int32)
 w = spmw.MemIn(int8) # resident weight
class ActIO(spmw.Interface):
 z_in = spmw.In(int32); y_out = spmw.Out(int8)

def mxu_links(i, j):
 return {
  WsIO.a_out: spmw.to((i, j+1), WsIO.a_in),
  WsIO.p_out: spmw.to((i+1, j), WsIO.p_in)}

@spmw.unit
def mac(io: WsIO):
 for m in range(Mt):
  a = io.a_in.get(); p = io.p_in.get()
  io.p_out.put(p + a * io.w); io.a_out.put(a)
\end{minted}
\end{minipage}\hfill
\begin{minipage}[t]{0.49\linewidth}
\vspace{0pt}
\begin{minted}[firstnumber=20]{python}
@spmw.fabric
def tpu_matmul(A: int8[Mt,Kt], W: int8[Kt,Nt],
               Y: int8[Mt,Nt]):
 T = spmw.Topology(WsIO, grid=(Kt, Nt),
                   link=mxu_links)
 P = spmw.place(mac, on=T)
 # act(io: ActIO) omitted for simplicity
 Ta = spmw.Topology(ActIO, grid=(Nt,))
 Pact = spmw.place(act, on=Ta)
 spmw.shard(W, into=P.w) # W[k,n]
 m = spmw.steps(Mt)
 spmw.stream_in(A, into=P.a_in,
                index=(m, P.rows))
 spmw.stream_in(0, into=P.p_in)
 spmw.stream_in(P.p_out, into=Pact.z_in)
 (lane,) = Pact.axes
 spmw.gather(Y, from_=Pact.y_out, index=(m, lane))
\end{minted}
\end{minipage}
\vspace{-6pt}
\caption{Weight-stationary matrix unit and activation row. Each MAC holds one weight, forwards activations, and adds to a partial sum. The activation body, constants, and imports are omitted.}
\label{fig:tpu}
\Description{The listing defines the weight-stationary MAC, the matrix and activation interfaces, and the fabric connecting matrix outputs to activation lanes. A comment at the activation instance group identifies the omitted body.}
\vspace{-10pt}
\end{figure}

\subsection{Weight-Stationary Systolic Array}
\label{sub:tpu}
First, we change a fixed-function array's dataflow with changes to PE computation and communication. \Name decouples boundary handling, so interior and edge PEs keep the same program.

\figref{fig:tpu} transforms the output-stationary GEMM in \figref[\subref*{subfig:spmw}]{fig:comparison} into a weight-stationary array. The original PE accumulates a result locally across $K$ products and forwards both operands. The new \texttt{mac} holds one weight, adds its product to an incoming partial sum, and forwards the activation. Accordingly, \texttt{WsIO} declares int8 activation streams, int32 partial-sum streams, and a scalar memory input for the weight. The connection rule sends activations horizontally and partial sums vertically. A separate \texttt{act} unit applies an activation function and fixed scaling to the result.

The fabric assigns $W[k,n]$ to PE $(k,n)$, supplies activations, seeds each column with zero, and connects the result to an activation lane. Derived boundaries let edge and interior PEs execute the same \texttt{mac}, receiving inputs from loaders or neighbors. This decouples dataflow changes from boundary handling and lets the activation unit evolve independently.

\subsection{Instruction-Programmable Mini-TPU}
\label{sub:tpuisa}
General neural network modules require reductions and elementwise operations beyond fixed matrix-and-activation computation. To support these operations without rebuilding hardware for each module, we add instruction control to the matrix and vector units. These units decode instructions locally, while the fabric supplies programs through the existing connections.

We port the matrix and vector operations of VTA's instruction set architecture (ISA)~\cite{moreau2018vta} into \texttt{@spmw.unit}, adapting instruction decoding and execution to our stream interfaces. Imperative \texttt{while} loops and \texttt{if}/\texttt{else} branches let a unit interpret run-time programs and update private state. Typed ports carry instructions and operands, while the fabric determines their delivery.

Programmable vector lanes replace the fixed activation row. Each receives int32 matrix results and a separate instruction stream. Its register-based program supports accumulation, arithmetic, shifts, maxima, and stores. The fabric binds matrix and vector programs independently and connects matrix outputs to vector inputs using the stream binding in \S~\ref{sub:streambindings}. Vector processing can change independently of the matrix array's connections. Array dimensions and storage capacities remain build parameters, while programs select operations without rebuilding hardware. \S~\ref{sub:programmable} evaluates LLaMA and DeepSeek feed-forward network (FFN) projections using this mini-TPU.

\begin{figure}[!t]
\setminted{fontsize=\fontsize{7pt}{8pt}\selectfont,
  xleftmargin=1.2em,numbersep=3pt}
\captionsetup[subfigure]{position=top,font=footnotesize,
  labelfont=bf,textfont=bf,skip=3pt}
\input{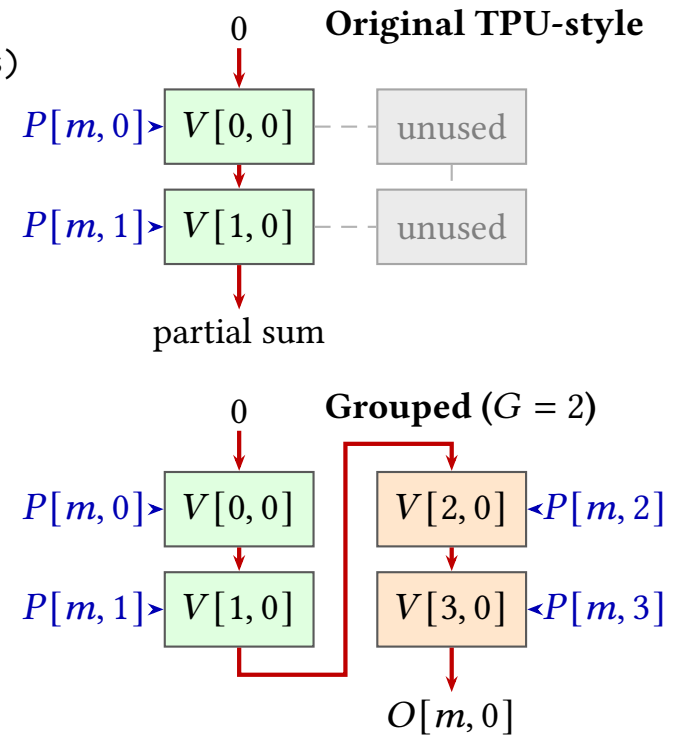}
\hfill
\begin{subfigure}[t]{0.32\linewidth}
\vspace{0pt}
\centering
\caption{Dataflow comparison}
\label{subfig:grouped-dataflow}
\begin{tikzpicture}[
 x=1pt,y=.82pt,font=\fontsize{7pt}{8pt}\selectfont,
 pe/.style={draw=black!65,fill=green!12,
   minimum width=28pt,minimum height=14pt,inner sep=1pt},
 idle/.style={pe,fill=black!8,draw=black!35,text=black!50},
 aarr/.style={-{Stealth[length=3pt]},blue!70!black,line width=.65pt},
 parr/.style={-{Stealth[length=3pt]},red!75!black,line width=.8pt},
 label/.style={inner sep=1pt},
 title/.style={font=\fontsize{7pt}{8pt}\selectfont\bfseries,inner sep=0pt}
]
\node[title,anchor=west] at (49,23) {Original TPU-style};
\foreach \k in {0,1} {
 \node[pe] (orig\k) at (33,-23*\k) {$V[\k,0]$};
 \node[idle] (idle\k) at (73,-23*\k) {unused};
 \node[label,blue!70!black] (op\k) at (4,-23*\k) {$P[m,\k]$};
 \draw[aarr] (op\k.east) -- (orig\k.west);
 \draw[black!30,densely dashed] (orig\k.east) -- (idle\k.west);
}
\draw[black!30,densely dashed] (idle0.south) -- (idle1.north);
\draw[parr] (33,18) node[above,black,label] {$0$} -- (orig0.north);
\draw[parr] (orig0.south) -- (orig1.north);
\draw[parr] (orig1.south) -- (33,-42)
 node[below,black,label] {partial sum};

\begin{scope}[yshift=-72pt]
\node[title,anchor=west] at (49,23) {Grouped ($G=2$)};
\foreach \k in {0,1} {
 \node[pe] (groupa\k) at (33,-23*\k) {$V[\k,0]$};
 \pgfmathtruncatemacro{\key}{\k+2}
 \node[pe,fill=orange!20] (groupb\k) at (73,-23*\k) {$V[\key,0]$};
 \node[label,blue!70!black] (ap\k) at (4,-23*\k) {$P[m,\k]$};
 \node[label,blue!70!black] (bp\k) at (102,-23*\k) {$P[m,\key]$};
 \draw[aarr] (ap\k.east) -- (groupa\k.west);
 \draw[aarr] (bp\k.west) -- (groupb\k.east);
}
\draw[parr] (33,18) node[above,black,label] {$0$} -- (groupa0.north);
\draw[parr] (groupa0.south) -- (groupa1.north);
\draw[parr] (groupa1.south) -- (33,-38) -- (53,-38)
 -- (53,15) -- (73,15) -- (groupb0.north);
\draw[parr] (groupb0.south) -- (groupb1.north);
\draw[parr] (groupb1.south) -- (73,-42)
 node[below,black,label] {$O[m,0]$};
\end{scope}
\end{tikzpicture}
\end{subfigure}
\caption{Grouped attention reuses the weight-stationary units in \figref{fig:tpu}. (a) Connections and fabric on a $(g,k,e)$ grid. (b) Two $2\times2$ arrays compute $P[4,4]V[4,1]$. Grouping joins both columns into one reduction chain.}
\label{fig:programming-examples}
\Description{Two panels show the group-link and attention-fabric code in two continuous columns beside two vertically stacked 2 by 2 arrays for P[4,4] times V[4,1]. In the original mapping above, the left column holds V[0,0] and V[1,0], and the right column is gray and unused. It produces a partial sum, then reloads the left column for a second pass. In the grouped mapping below, the green left column and orange right column are both active, holding V[0,0] through V[3,0]. A red partial-sum path starts at zero, descends the left column, connects to the top of the right column, and descends to O[m,0]. Blue arrows supply each PE with its corresponding P value.}
\vspace{-18pt}
\end{figure}

\subsection{Grouped Attention-PV Array}
\label{sub:attnpv}
Beyond prototyping existing designs, hardware engineers need to explore new architectures for emerging workloads. Instruction control cannot reconnect a fixed TPU-style datapath to match operand shapes. For example, during large language model (LLM) prefill~\cite{pope2023scalellmtpu}, a head with sequence length $S$ and head dimension $d_h$ has $Q$, $K$, and $V$ of shape $[S,d_h]$. Scaling, masking, and softmax of $QK^\top$ produce probabilities $P[S,S]$. Attention-PV then computes $O[S,d_h]=P[S,S]V[S,d_h]$. Modern TPUs like TPUv7 have $256\times256$ systolic matrix units (MXUs) that hold and reuse the right-hand operand~\cite{jiang2026ragged}, while models like LLaMA~3~8B have head dimension $d_h=128$~\cite{dubey2024llama3}. This limits the reduction dimension in $QK^\top$ and the output width in $PV$ to 128. Both attention matmuls therefore achieve at most 50\% MXU utilization under this per-head mapping~\cite{jiang2026ragged}.

One architectural change to increase utilization is to use idle columns for the same head's $PV$ reduction, reusing the units in \figref{fig:tpu}. For an $R\times C$ array with $d_h<C$ dividing $C$, we split the columns into $G=C/d_h$ groups, forming a $G\times R\times d_h$ grid indexed by group $g$, row $k$, and lane $e$. In each $GR$-position reduction tile, group $g$ holds $V$ rows $gR$ through $(g+1)R-1$ and receives matching $P$ columns. Activations stay within groups, while partial sums descend a column and continue at the next group's top, forming $d_h$ chains of depth $GR$. In \figref[\subref*{subfig:grouped-dataflow}]{fig:programming-examples}, the original $2\times2$ mapping uses two PEs (50\%) in two passes for $d_h=1$. Grouping uses all four PEs (100\%) in one pass. With $R=C=256$ and $d_h=128$, two groups process 512 sequence positions per pass.

\figref[\subref*{subfig:attn-fabric}]{fig:programming-examples} implements these connections and bindings on tiles with \texttt{Mt} query positions and \texttt{Lt} $=GR$ reduction positions. The connection rule joins groups, while the binding maps select tile elements $V[gR+k,e]$ and $P[m,gR+k]$. Derived boundaries expose the new activation inputs, seed only the first group, and drain only the last. Adding a group requires no per-PE edits or separate enumeration of loaders and drains. Partial sums accumulate across reduction tiles before activation.

\Name enables rapid hardware--software co-design by decoupling arithmetic units from topology and data delivery. Designers can explore new connections and workload mappings through shared declarations while reusing unit implementations. \S~\ref{sub:reconfig} evaluates the resulting utilization and pipeline-depth tradeoffs at equal PE counts.

\section{Compilation Flow}
\label{sec:compile}

\begin{figure}[t]
\centering
\begin{tikzpicture}[
  x=1pt,y=1pt,font=\fontsize{8.5pt}{9.5pt}\selectfont,
  box/.style={draw=black!65,rounded corners=2pt,align=center,
              inner sep=3pt,minimum height=19pt,fill=black!5},
  arr/.style={-{Stealth[length=3pt]},semithick},
  feedback/.style={dashed,semithick,draw=black!60}
]
\node[box,text width=34pt] (source) at (20,0) {SPMW design};
\node[box,text width=43pt] (frontend) at (74,0) {Python frontend};
\node[box,text width=58pt,fill=orange!15] (dialect) at (138,0)
  {SPMW dialect\\(MLIR)};
\node[box,text width=58pt,fill=orange!15] (functional) at (138,-36)
  {Functional\\simulation};
\node[box,text width=39pt,fill=blue!10] (cpp) at (213,13) {HLS C++};
\node[box,text width=54pt,fill=green!12] (rtl) at (213,-13) {Structural RTL};
\node[box,text width=42pt,fill=violet!10] (integrate) at (292,0) {RTL\\integration};
\node[box,text width=52pt] (place) at (361,0) {Placement\\\& routing};
\node[box,text width=42pt,fill=violet!10] (rtlsim) at (292,-36) {RTL\\simulation};
\node[box,text width=52pt] (build) at (361,-36) {Binary build};
\draw[arr] (source) -- (frontend);
\draw[arr] (frontend) -- (dialect);
\draw[arr] (dialect.south) -- (functional.north);
\draw[arr] (dialect.east) -- ++(6,0) |- (cpp.west);
\draw[arr] (dialect.east) -- ++(6,0) |- (rtl.west);
\draw[arr] (cpp.east) -- (258,13) |- ([yshift=5pt]integrate.west);
\node[font=\fontsize{7.5pt}{8pt}\selectfont,above,inner sep=1pt]
  at (251,15) {HLS tool};
\draw[arr] (rtl.east) -- (258,-13) |- ([yshift=-5pt]integrate.west);
\draw[arr] (integrate.east) -- (place.west);
\draw[arr] (integrate.south) -- (rtlsim.north);
\draw[arr] (place.south) -- (build.north);
\draw[feedback] (functional.south) -- (138,-55);
\draw[feedback] (rtlsim.south) -- (292,-55);
\draw[arr,feedback] (build.south) -- (361,-55) -- (20,-55) -- (source.south);
\end{tikzpicture}
\caption{Compilation and simulation flow. Unit bodies and memory movers pass through HLS (blue), while fabric structure drives RTL assembly (green). Dashed paths show feedback for iterative design refinement.}
\label{fig:overview}
\Description{The Python frontend lowers an SPMW design into the SPMW MLIR dialect, which branches to functional simulation. The blue branch generates HLS C++ for unit bodies and memory movers and passes it through an HLS tool. The green branch generates structural RTL from fabric instance groups, links, and bindings. RTL integration joins both hardware paths and branches to RTL simulation. A separate path proceeds from integration through placement and routing to binary generation. Dashed paths from functional simulation, RTL simulation, and binary generation merge into a return arrow to the SPMW design, showing iterative refinement.}
\end{figure}
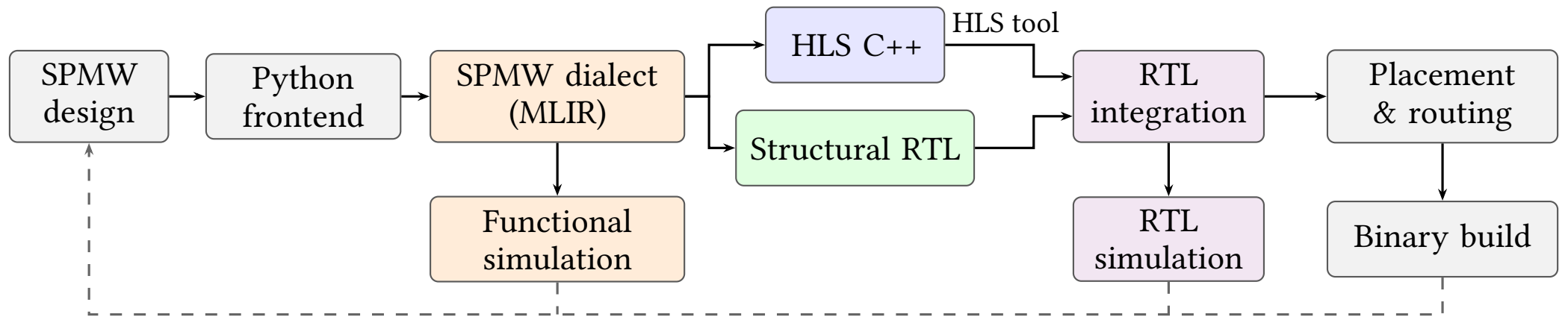

Flattening a regular accelerator into independent kernels can duplicate scheduling work and obscure the relationship between PE behavior and array connectivity. \Name preserves this relationship so the compiler can specialize shared PE code for reuse in a fabric generated from the same architectural description. In \figref{fig:overview}, our Python frontend lowers \Name directly to MLIR. The \texttt{@spmw.unit} bodies become HLS C++ for scheduling and pipelining. Elaborating \texttt{@spmw.fabric} determines how these kernels connect and receive data in the assembled RTL.

\subsection{Structural Elaboration and Functional Simulation}
\label{sub:elaboration}
We design an \Name MLIR dialect to preserve regularity through compilation. Each instance group retains a logical grid and shared unit symbol, with rules for communication and boundary supply. The frontend parses each unit's Python abstract syntax tree (AST), resolves port accesses and grid coordinates against the fabric, and performs type inference to emit MLIR functions with arithmetic, control flow, and memory operations.

Elaborating connection rules derives internal channels and exposed boundaries, which the compiler checks against interfaces and fabric bindings. Port usage and static parameters determine kernel specialization. Shared symbols keep instances associated with their unit function, so topology changes update communication and specialization while compatible instances reuse the function. The compiler therefore works directly from the generating description.

Within an instance group, all instances share a unit definition, static design parameters, and synthesis settings. For each coordinate $x$ in the logical grid, the compiler constructs the following specialization key
\[
  \chi(x)=(s_x,r_x,c_x),\qquad
  x\sim y\ \Longleftrightarrow\ \chi(x)=\chi(y).
\]
Here, $s_x$ is the bound-port signature selecting the unit body, $r_x$ records how its port accesses are routed in generated code, and $c_x$ contains statically specialized coordinate values. The compiler resolves these keys from the topology and bindings, partitions the grid by key equality, and lowers one representative per class. It specializes the body and interface and synthesizes them once. RTL assembly instantiates the module at each class member with independent state and connections.

Furthermore, we implement a parallel functional simulator that reuses unit functions across the grid before hardware synthesis. Unit instances and memory movers become MLIR calls in \texttt{omp.section} blocks within an \texttt{omp.parallel} region. One thread per section keeps producers and consumers scheduled when a task blocks. Each typed channel connects one producer's stream output to one consumer's input through a bounded FIFO with synchronized access. Capacity uses the largest endpoint \texttt{depth} request, or two values by default. The simulator uses at least one slot per channel and otherwise preserves these capacities by default. The compiler lowers the MLIR to LLVM for host execution with OpenMP. Comparing outputs with software references provides quick correctness feedback, but we currently do not support performance-oriented simulation (\S~\ref{sec:discussion}).

\subsection{HLS Kernel Generation}
\label{sub:dialect}
HLS compiles unit computation, including instruction decoding, and generated memory movers. The flow also supports traditional loop-based generation by expressing an entire loop nest within a single \texttt{spmw.unit} for HLS synthesis. Instances share an implementation only when their specialized kernel definitions and synthesis settings match. Specialization uses bindings to remove unused forwarding outputs. Coordinate-dependent constants and port usage can require different configurations, so sharing a unit symbol is insufficient. The compiler generates one HLS C++ kernel per distinct configuration and instantiates its IP with independent state for each PE.

Scheduling each distinct kernel once avoids repeating expensive HLS optimization across identical instances. For example, enlarging a systolic grid adds interior PEs without necessarily adding kernel configurations. Edge PEs can require separate configurations when their forwarding ports differ. HLS work therefore depends on the distinct specialized kernels, while the RTL retains the full spatial array. Independent kernel projects can also run concurrently within the worker budget.

\begin{figure}[t]
\centering
\includegraphics[width=\columnwidth]{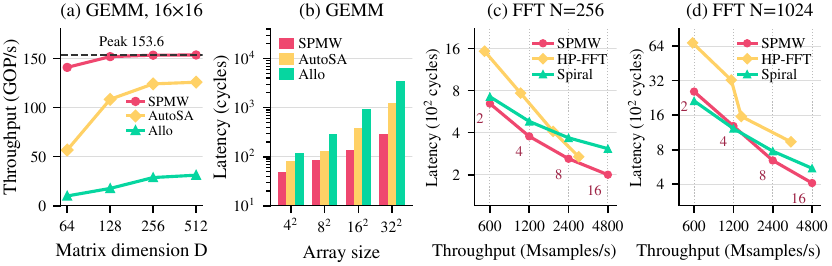}
\caption{(a) $16\times16$ GEMM throughput ($M=N=K=D$), with theoretical peak 153.6~GOP/s dashed. (b) GEMM latency ($M=N=K=D$) on $D\times D$ arrays. (c,d) FFT latency (cycles) versus sustained throughput for $N=256,1024$. \Name point labels give samples/cycle. Routed frequencies are capped at 300~MHz.}
\label{fig:single-kernel}
\Description{Four equally wide panels. The first compares whole-GEMM throughput on 16-by-16 arrays at matrix dimensions 64, 128, 256, and 512. Frequencies are capped at 300 MHz. SPMW reaches 140.94, 151.89, 153.38, and 153.57 GOP/s, approaching the dashed theoretical peak of 153.6 GOP/s. AutoSA reaches 56.71 to 125.83 GOP/s and Allo 9.95 to 31.14 GOP/s. Panel (b) reports latency in cycles. Panels (c) and (d) label latency in hundreds of cycles, with ticks 2, 4, 8, 16 and 4, 8, 16, 32, 64, respectively. Each FFT panel has independent logarithmic y-axis limits and labeled ticks, with ranges 120 to 2200 cycles at N=256 and 260 to 9000 cycles at N=1024. The GEMM bars compare SPMW, AutoSA, and Allo on complete M=N=K=D workloads using D-by-D arrays. The third and fourth panels plot FFT completion latency against sustained streaming throughput in millions of complex samples per second for 256 and 1024 points, respectively, with vertical grid lines. All panels use the same coral, gold, and mint palette. In each FFT panel, colors and circle, diamond, and triangle markers identify SPMW, HP-FFT, and Spiral. Both FFT panels include legends and label SPMW points with their widths below the curve. Each curve has four points, corresponding from left to right to widths 2, 4, 8, and 16 complex samples per cycle. Throughput is N times the capped clock frequency divided by the measured steady transform interval.}
\end{figure}

\subsection{RTL Integration}
\label{sub:lowering}\label{sub:rtl}
From the elaborated fabric, the RTL generator instantiates specialized kernel IP, materializes topology links as FIFO channels, and exposes residual ports as boundary bundles. Repeated PEs share a module definition but retain independent state and connections.

Vitis HLS exports each kernel IP~\cite{amd2023hlsexport}. Generated wrappers map its signal names to fabric ports. For each channel, structural RTL connects the producer and consumer through payload signals, read/write enables, and empty/full feedback~\cite{amd2023hlsinterfaces}. Boundary bindings supply constant sources or connect external streams and supported memory movers. For memory-connected kernels, the generator assembles loaders, drains, and command control around the PE array. This path realizes data delivery directly from the fabric description without a monolithic HLS schedule for the array.

Vivado synthesizes the assembled RTL, and selected designs go through placement, routing, and bitstream generation. RTL assembly, logic synthesis, and routing still process the instantiated design. As kernel synthesis is amortized, these stages can dominate total build time.

\section{Evaluation}
\label{sec:exp}
In this section, we evaluate \Name's performance, resource usage, compilation and simulation time, and design effort. We use systolic GEMM to test replication with nearest-neighbor communication and FFT to test butterfly connections that change across stages. To evaluate composed accelerators, we implement an instruction-programmable mini-TPU and a research accelerator FEATHER with programmable reduction and reordering~\cite{tong2024feather}, and compare them with baseline implementations.

\textbf{AI Use.} \ifdefined\SPMWArxiv We build on top of Allo~\cite{chen2024allo} and reuse the compiler skeleton\else We manually implement a compiler skeleton\fi{} that lowers Python ASTs to standard MLIR dialects and emits C++ HLS. After finalizing the \Name specification in \S~\ref{sec:model}, we use Claude Code with Opus~5 to add frontend extensions, MLIR lowering, and RTL glue logic\ifdefined\SPMWArxiv\else{} to this skeleton\fi.

\textbf{Experimental setup.}
\label{sub:setting}
We target the AMD Alveo U280 FPGA~\cite{u280} at a default 300~MHz and leave ASIC evaluation to future work. In the following, we use Vitis HLS and Vivado 2023.2~\cite{vitis_hls_2023,vivado_2023}. All experiments run on a server with two Intel Xeon Silver 4214 processors, totaling 24 physical cores and 48 hardware threads. We measure RTL cosimulation latency in cycles and run placement and routing (P\&R) to obtain resource counts and timing.
\reviewissue{R1}{E1 latency remains pinned at 129f474. SPMW resources use the full-kernel correction at 13ab1d2, including memory movers and the control interface, checked against the raw routed reports. The fixed-16x16 whole-GEMM sweep in panel (a) uses E12 at 13ab1d2 (12 October), selecting kernel scope and the corrected AutoSA host-serialize/K-tiled/latency[4,4] configuration. All twelve points pass RTL checks. AutoSA and Allo at 512 use start-to-done cycles because their memory-beat waveforms were not retained. Other points use first-to-last memory beats. Throughput uses min(300 MHz, unrounded routed fmax). Incomplete 1024-cubed simulations are omitted. FFT E2/E11 use b98646c (11 October), including Spiral and the N=1024 width sweep. E4 row-loader P\&R and E5 are checked at 23eedb2 (12 September). Read the current CSV variants rather than stale aggregate JSON. E5's source-to-RTL comparison now uses 31673105, with measured one-worker SPMW runs and frontend timings for the legacy baselines. E4 convolution is updated from 1ef4599 with all six row-loader rows checked against raw reports. Mini-TPU \figref{fig:mini-tpu} and complete-engine resources use 98c8486 (3 October): measured LLaMA and DeepSeek gate/up slices across array sizes 4/8/16, with effective throughput derived from whole-engine cycles and the best reported routed frequency capped at 300 MHz. Earlier normalization and hybrid board results remain archived. E6 and agent records are unchanged. LoC retains the 23eedb2 counts for unchanged designs, restores the archived folded FFT, and adds a fresh FEATHER RTL count.}

\subsection{Performance and Resource Usage}
\label{sub:perf}

\subsubsection{Compute and Connectivity}
\label{sub:compute}
We first evaluate GEMM and FFT performance and resources across different communication topologies.

\begin{table}[t]
\centering
\caption{U280 performance and routed resources at $f=\min(300,f_{\mathrm{routed}})$ MHz. GEMM throughput is $2D^3f/(1000L)$ GOP/s for dimension $D=M=N=K$ and latency $L$ cycles. For $N$-point FFTs, throughput is $Nf/I_{\mathrm{FFT}}$ MS/s, with steady interval $I_{\mathrm{FFT}}$ cycles and width $w$ samples/cycle.}
\label{tab:resources}
\footnotesize
\setlength{\tabcolsep}{2pt}
\renewcommand{\arraystretch}{0.95}
\begin{tabular}{@{}llrrrrrrr@{}}
\toprule
\textbf{Design} & \textbf{Config.} & \makecell{\textbf{Latency}\\\textbf{(cycles)}} & \makecell{\textbf{Throughput}\\\textbf{(GOP/s or MS/s)}} & \textbf{LUT} & \textbf{FF} & \textbf{DSP} & \textbf{BRAM18} & \makecell{\textbf{Freq.}\\\textbf{(MHz)}} \\
\midrule
\multicolumn{9}{l}{GEMM (int8/int32, $M=N=K=32$)} \\
\Name & $32^2$ & \textbf{280} & \textbf{70.22} & 147,551 & 193,953 & 1,024 & 46 & 330 \\
AutoSA & $32^2$ & 1,222 & 16.09 & 186,369 & 320,772 & 1,024 & 48 & 309 \\
Allo & $32^2$ & 3,408 & 4.82 & 151,663 & 109,251 & 1,024 & 3 & 251 \\
\midrule
\multicolumn{9}{l}{FFT (complex FP32, 1,024 points)} \\
\Name & $w=16$ & \textbf{410} & \textbf{4,800.00} & 220,111 & 257,729 & 678 & 96 & 306 \\
HP-FFT & $w=16$ & 937 & 3,308.90 & 263,862 & 177,909 & 666 & 404 & 239 \\
Spiral & $w=16$ & 551 & \textbf{4,800.00} & 172,301 & 255,005 & 440 & 336 & 327 \\
\bottomrule
\end{tabular}
\Description{Latency, throughput, LUT, flip-flop, DSP, BRAM18, and frequency for complete M=N=K=32 GEMMs on 32-by-32 arrays using SPMW, AutoSA and Allo, and 1024-point streaming FFTs using SPMW, HP-FFT and Spiral. GEMM throughput uses operations divided by single-workload latency. FFT throughput uses the measured steady transform interval. Best latency and throughput are bold.}
\vspace{-10pt}
\end{table}

\textbf{Output-stationary systolic GEMM.}
We compare int8 multiplication with int32 accumulation in \Name against two accelerator design frameworks. AutoSA~\cite{wang2021autosa} is a state-of-the-art polyhedral systolic-array generator that derives PE computation and I/O networks from loop nests, with default Vitis HLS and TAPA~\cite{guo2023tapa} backends. We use Allo's systolic-array library~\cite{chen2024allo}, which unfolds PE loops and uses communication primitives to stream data between PEs.

\figref[a]{fig:single-kernel} measures $64^3$--$512^3$ GEMMs on a $16\times16$ array. \Name's boundary bindings generate separate A and B feed chains that connect 512-bit memory interfaces directly to edge FIFOs, overlapping delivery and computation. It drains sums while PEs compute the next tile, eliminating boundary stalls and reaching 91.76\%--99.98\% of peak. AutoSA also overlaps input delivery through double buffering, but adds distribution and aggregation stages. Its serialized result chain delays short workloads. It reaches 99.2\% cycle efficiency at $512^3$, but routing from FIFO control to accumulator BRAM limits its clock to 248~MHz, versus 300~MHz for \Name. Allo's library implementation copies all of A and then B into local arrays, executes the tiles, and finally writes C. Its generated C++ has no outer HLS dataflow region across these stages, so PE-to-PE streaming does not overlap memory transfers with computation. Its 32-bit ports further increase transfer time, but widening them alone would not remove this sequential schedule. Overlap would require restructuring its load, compute, and store stages and their buffers. \Name combines high utilization with workload flexibility: all points use one hardware build, with the reduction length and tile count sent before the operands on each row's input stream. Boundary units use these values to mark tile ends for PEs. The AutoSA and Allo designs fix dimensions at generation time and require a separate build per workload.

To better understand \Name's scalability, we measure complete $M=N=K=D$ workloads on $D\times D$ arrays, ranging from $4\times4$ to $32\times32$ in \figref[b]{fig:single-kernel}. Latency spans the first input-memory transfer through the final result write, excluding host communication and control. \Name reduces latency by $1.59$--$4.36\times$ over AutoSA and $2.51$--$12.17\times$ over Allo, with the largest gains at $32\times32$. All three use one DSP per PE and a one-cycle reduction-loop initiation interval (II), so \Name's gains come from shorter delivery and drain paths and overlapped transfers. Table~\ref{tab:resources} reports $32\times32$ performance and resources. \Name uses fewer LUTs and FFs than AutoSA with similar BRAM usage. Both meet 300~MHz. Allo misses because its result-store control reaches an array buffer through a high-fanout net dominated by routing delay.

\reviewissue{R2}{Keep one AutoSA series selecting the lowest measured latency at each size. E1 latency uses the first input-memory transfer through the final result write, including the completed Allo S16/S32 records. Native-call latency remains in the source evidence, not a second latency definition in the paper. The interval column is removed at the author's request. Allo S32 repeated-call RTL interval evidence (3427--3458 cycles, HLS estimate 3273) remains archived in the 13 October interval audit. GEMM throughput remains operations divided by memory-transfer latency. Checked 17 October against upstream 2366056: E1 and E12 CSVs and E1 generation scripts match the plotted snapshots. Panel (b) uses array\_part[S,S,S], latency[1,1], simd[1], and host serialization at sizes 8, 16, 32. Size 4 remains narrow because the serializer fails on the sub-beat operand. Panel (a) uses host serialization, array\_part[64,64,Kt], latency[4,4], simd[1], with Kt=32 at 64 cubed and 64 otherwise. These are different workload configurations. The checked E12 Allo C++ has sequential load\_buf0, load\_buf1, systolic\_gemm, and store\_res2 calls, and sequential tile loads, compute, and store. No outer dataflow region overlaps those stages.}
\reviewissue{R3}{Allo S32 values remain LUT 151663, FF 109251, DSP 1024, WNS -0.657 ns. The latest E1 CSV has corrected BRAM18 units: AutoSA S32 wide has 48 and Allo has 3. Do not multiply this corrected column by two again. Raw Allo utilisation and timing reports are checked. The E1 array-only Gemmini comparison remains removed.}

\textbf{Radix-2 FFT.}
\label{sub:fft}
We compare complex FP32 transforms with HP-FFT~\cite{wang2025hpfft}, a state-of-the-art manually optimized HLS implementation, and Spiral~\cite{milder2012spiral}, an RTL FFT generator. \figref[c,d]{fig:single-kernel} plots latency against sustained streaming throughput for 256- and 1,024-point designs at $w=2,4,8,16$ samples/cycle. HP-FFT uses $UF=w/2$ butterfly lanes per stage. Spiral uses fully streaming radix-2 cores with natural input/output order. Latency spans the first accepted input through the final output of one transform.

\begin{figure}[t]
\centering
\begin{minipage}[c]{0.52\linewidth}
\includegraphics[width=\linewidth]{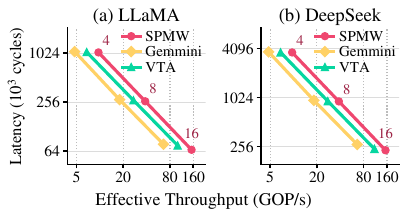}
\end{minipage}\hfill
\begin{minipage}[c]{0.47\linewidth}
\begingroup
\centering
{\small\bfseries (c) Resources, $16\times16$ engines}\par\smallskip
\footnotesize
\setlength{\tabcolsep}{1.5pt}
\begin{tabular}{@{}lrrrrrr@{}}
\toprule
\textbf{Design} & \textbf{LUT} & \textbf{FF} & \textbf{DSP} & \makecell{\textbf{BRAM}\\\textbf{(18Kb)}} & \textbf{URAM} & \makecell{\textbf{Freq.}\\\textbf{(MHz)}} \\
\midrule
\Name & 35,544 & 33,603 & 0 & 28 & 0 & 334 \\
Gemmini & 65,529 & 39,989 & 29 & 32 & 8 & 153 \\
VTA & 33,923 & 7,387 & 0 & 140 & 12 & 224 \\
\bottomrule
\end{tabular}
\par\endgroup

\end{minipage}
\caption{Programmable engines on 64-token FFN gate/up slices. (a,b) Cycle latency versus effective throughput at routed frequency capped at 300~MHz. Labels give array side length. (c) $16\times16$ routed resources.}
\label{fig:mini-tpu}
\Description{Two panels plot latency in thousands of cycles against effective throughput in GOP/s for LLaMA 3.2 1B and a DeepSeek-V4-Pro expert. Each curve connects array sizes 4-by-4, 8-by-8, and 16-by-16, with point labels giving the side length. Coral circles identify SPMW, gold diamonds Gemmini, and mint triangles VTA. Lower and farther right is better. Throughput counts two operations per MAC and uses routed frequency capped at 300 MHz. SPMW, Gemmini matmul-only, and VTA include instruction delivery, operand reads and result writes. The resource panel gives LUT, FF, DSP, BRAM18, URAM and MHz at size 16. SPMW uses 35544 LUTs, 33603 FFs, 0 DSPs, 28 BRAM18, 0 URAM and a routed frequency of 334 MHz, capped at 300 MHz for throughput.}
\vspace{-10pt}
\end{figure}

Relative to HP-FFT, \Name reduces latency by $1.35$--$2.37\times$ at 256 points and $2.29$--$2.65\times$ at 1,024 points. HP-FFT's transform-sized ping-pong buffers delay each consumer until its producer completes the transform. \Name's explicit stage connections lower to scalar FIFOs, letting stages overlap within a transform as operands arrive. HP-FFT's inner loops achieve II=1, but its stage processes drain and restart between transforms in our Vitis 2023.2 toolchain, adding roughly 10--13 cycles beyond the ideal $N/w$ interval. Both \Name and Spiral achieve $N/w$ at every measured width, so their latency difference comes from pipeline startup. \Name's coordinate-based rules specialize butterfly connections by stage and lane. Permutation units buffer partners arriving on different beats, and a final reorder buffers one transform for natural output order. These fill delays grow with $N/w$, leaving \Name slower than Spiral at 1,024 points and $w=2,4$. Wider lanes shorten the delays and let more stages connect simultaneous partners directly, without permutation buffers. This reduces \Name's startup enough to outperform Spiral at $w=8,16$. At 256 points, the shorter buffers give \Name lower latency at every width.

Table~\ref{tab:resources} reports the widest 1,024-point FFT configurations. HP-FFT's partitioned ping-pong arrays duplicate many small BRAM banks. \Name uses short FIFOs and local register delays, reserving BRAM for twiddle tables. Its lower BRAM usage comes with more FFs than HP-FFT. All measured \Name and Spiral configurations meet 300~MHz. HP-FFT misses at several wide configurations. In the failing 1,024-point configurations, the critical paths pass through FIFO and control signals in the bit-reversal stage, with about 85\% of their delay from routing. Spiral uses fewer LUTs and DSPs and achieves higher routed frequency, but matches \Name's throughput under the 300~MHz cap. Moreover, Spiral rounds toward zero and uses two DSPs per multiplier, while the HLS designs use Vitis floating-point units with three. This contributes to \Name's higher DSP count.
\reviewissue{R11}{FFT uses commit b98646c: E2 at N=256 and E11 at N=1024. All widths use SPMW lanes and the corrected HP-FFT UF4/UF8 configurations. Spiral uses the copied SwizzleFFT 6d1a2af simulation and routed reports. Eight Spiral transforms and 32 SPMW/HP-FFT transforms are checked (33 for HP-FFT UF1). The CSV and available raw reports are archived with Git blob hashes. E2 wide-lanes raw reports remain absent upstream. The Allo FFT performance comparison is removed. The separate Code Size comparison retains the author-selected folded SPMW/HP-FFT source counts.}

\subsubsection{Composability}
We use mini-TPU and FEATHER to demonstrate how \Name composes different hardware components into complete accelerators.

\textbf{Mini-TPU.}
\label{sub:programmable}
We compare \Name with Gemmini~\cite{genc2021gemmini}, a systolic DNN accelerator generator, and VTA~\cite{moreau2018vta}, a programmable tensor accelerator integrated with TVM~\cite{chen2018tvm}. These engines combine matrix computation, runtime control, and vector processing. \Name composes its systolic array with instruction dispatch, vector lanes, and memory movers through typed interfaces. One GEMM instruction selects tensor addresses, tile dimensions, reduction length, and the epilogue.

\begin{figure}[t]
\begin{minipage}[t]{0.61\linewidth}
\vspace{0pt}
\centering
\begin{tikzpicture}[
  font=\fontsize{7pt}{8pt}\selectfont,
  sw/.style={circle, draw=black!75, fill=yellow!22,
    minimum size=0.40cm, inner sep=0pt, font=\bfseries\fontsize{7pt}{8pt}\selectfont},
  edge/.style={-{Stealth[length=2.5pt]}, black!75, line width=0.4pt},
  leftflow/.style={-{Stealth[length=3pt]}, green!45!black, line width=0.85pt},
  rightflow/.style={-{Stealth[length=3pt]}, blue!70!black, line width=0.85pt},
  sumflow/.style={-{Stealth[length=3pt]}, orange!85!black, line width=0.85pt},
  head/.style={anchor=west, font=\bfseries\fontsize{7.5pt}{8.5pt}\selectfont}
]
\node[head] at (0.02,0.48) {(a) BIRRD, $AW=8$};
\node[head] at (4.77,0.48) {(b) Switch $S$};
\node[draw=black!75, fill=black!3, minimum width=3.65cm,
      minimum height=0.36cm, inner sep=1pt] at (2.30,0)
  {NEST PE array ($AH\times8$)};
\fill[blue!3] (0.46,-0.32) rectangle (4.15,-3.57);

\foreach \s in {0,1,2,3,4,5} {
  \node[anchor=east, font=\fontsize{6pt}{7pt}\selectfont]
    at (0.42,-0.61-0.54*\s) {$s=\s$};
  \foreach \j in {0,1,2,3} {
    \node[sw] (b\s\j) at (0.80+\j,-0.61-0.54*\s) {$S$};
  }
}
\foreach \s/\bits in {0/2,1/3,2/3,3/3,4/2} {
  \foreach \q in {0,...,7} {
    \ifnum\bits=2
      \pgfmathtruncatemacro{\dest}{{0,2,1,3,4,6,5,7}[\q]}
    \else
      \pgfmathtruncatemacro{\dest}{{0,4,2,6,1,5,3,7}[\q]}
    \fi
    \pgfmathsetmacro{\srcx}{0.80+floor(\q/2)+(2*mod(\q,2)-1)*0.12}
    \pgfmathsetmacro{\dstx}{0.80+floor(\dest/2)+(2*mod(\dest,2)-1)*0.12}
    \draw[edge] (\srcx,-0.77-0.54*\s) -- (\dstx,-0.99-0.54*\s);
  }
}
\foreach \q in {0,...,7} {
  \pgfmathsetmacro{\x}{0.80+floor(\q/2)+(2*mod(\q,2)-1)*0.12}
  \draw[edge] (\x,-0.18) -- (\x,-0.45);
  \draw[edge] (\x,-3.47) -- (\x,-3.64);
}
\node[draw=black!75, fill=black!3, minimum width=3.65cm,
      minimum height=0.36cm, inner sep=1pt] at (2.30,-3.82)
  {Output buffers};

\node[sw, draw=orange!85!black, fill=orange!22] (focus) at (b53) {$S$};
\draw[orange!85!black, dashed, line width=0.7pt]
  (focus.north east) -- (4.68,0.08);
\draw[orange!85!black, dashed, line width=0.7pt]
  (focus.south east) -- (4.68,-3.81);
\draw[orange!85!black, dashed, rounded corners=2pt, line width=0.7pt]
  (4.68,0.08) rectangle (8.14,-3.81);
\node[orange!75!black] at (6.45,-3.55) {2-bit command per switch};

\foreach \x/\y/\mode in {5.55/-0.80/0,7.35/-0.80/3,5.55/-2.43/1,7.35/-2.43/2} {
  \begin{scope}[shift={(\x,\y)}]
    \draw[black!65, fill=white, line width=0.5pt] (0,0) circle (0.46cm);
    \node[green!40!black] at (-0.27,0.65) {$L$};
    \node[blue!70!black] at (0.27,0.65) {$R$};
    \draw[leftflow] (-0.27,0.52) -- (-0.27,0.32);
    \draw[rightflow] (0.27,0.52) -- (0.27,0.32);
    \ifnum\mode=0
      \draw[leftflow] (-0.27,0.32) -- (-0.27,-0.52);
      \draw[rightflow] (0.27,0.32) -- (0.27,-0.52);
      \node at (0,-0.78) {pass (00)};
    \fi
    \ifnum\mode=3
      \draw[leftflow] (-0.27,0.32) -- (0.27,-0.32) -- (0.27,-0.52);
      \draw[rightflow, preaction={draw=white,line width=2pt}]
        (0.27,0.32) -- (-0.27,-0.32) -- (-0.27,-0.52);
      \node at (0,-0.78) {swap (11)};
    \fi
    \ifnum\mode=1
      \node[draw=orange!85!black, fill=orange!18, inner sep=1pt]
        (plus) at (0.20,-0.02) {$+$};
      \draw[leftflow] (-0.27,0.32) -- (-0.27,-0.52);
      \draw[leftflow] (-0.27,0.20) -- (plus.west);
      \draw[rightflow] (0.27,0.32) -- (plus.north);
      \draw[sumflow] (plus.south) -- (0.27,-0.52);
      \node at (0,-0.78) {add-right (01)};
    \fi
    \ifnum\mode=2
      \node[draw=orange!85!black, fill=orange!18, inner sep=1pt]
        (plus) at (-0.20,-0.02) {$+$};
      \draw[rightflow] (0.27,0.32) -- (0.27,-0.52);
      \draw[rightflow] (0.27,0.20) -- (plus.east);
      \draw[leftflow] (-0.27,0.32) -- (plus.north);
      \draw[sumflow] (plus.south) -- (-0.27,-0.52);
      \node at (0,-0.78) {add-left (10)};
    \fi
  \end{scope}
}
\end{tikzpicture}
\end{minipage}\hfill
\begin{minipage}[t]{0.375\linewidth}
\vspace{0pt}
\textbf{\scriptsize (c) Topology and boundary composition}\par\smallskip
\begin{minted}[fontsize=\fontsize{6.5pt}{7.5pt}\selectfont]{python}
AW = 8; LG = int(log2(AW)); P0 = 2*LG
def wiring(s, j):
 if s == P0-1: return {}
 bits = (2 if s == 0
         else min(LG, 2+s, 2*LG-s))
 links = {}
 for port,q in ((SwIO.out_l,2*j),
                (SwIO.out_r,2*j+1)):
  pos = reverse_bits(q, bits)
  dst = (SwIO.in_l if pos % 2 == 0
         else SwIO.in_r)
  links[port] = spmw.to((s+1, pos//2), dst)
 return links
birrd = spmw.Topology(SwIO,
    grid=(P0, AW//2), link=wiring)
# In the fabric, P holds the NEST grid.
B = spmw.place(switch, on=birrd)
spmw.stream_in(P.p0_out, into=B.in_l)
spmw.stream_in(P.p1_out, into=B.in_r)
\end{minted}
\end{minipage}
\vspace{-5pt}
\caption{BIRRD at \mbox{$AW=8$}. (a, b) Six stages of four two-input, two-output switches. (c) The topology reverses low lane-index bits and binds NEST to BIRRD. Unit bodies, interfaces, and \texttt{reverse\_bits} are omitted.}
\label{fig:feather-topology}
\Description{NEST feeds eight lanes into six rows of four BIRRD switches and then into output buffers. The five interstage permutations reverse two, three, three, three, and two low lane-index bits. An inset shows a switch's four operations: pass 00 yields L,R. Swap 11 yields R,L. Add-right 01 yields L,L+R. Add-left 10 yields L+R,R. The code generates these links and binds NEST's paired partial-sum outputs to BIRRD's left and right inputs.}
\end{figure}

\begin{figure}[t]
\centering
\begin{minipage}[c]{0.55\linewidth}
\includegraphics[width=\linewidth]{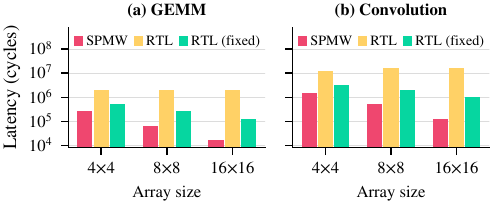}
\end{minipage}\hfill
\begin{minipage}[c]{0.435\linewidth}
\centering
\begingroup
\centering
{\small\bfseries (c) Resources}\par\smallskip
\footnotesize
\setlength{\tabcolsep}{2pt}
\renewcommand{\arraystretch}{0.9}
\begin{tabular}{@{}llrrr@{}}
\toprule
\textbf{Size} & \textbf{Design} & \textbf{LUT} & \textbf{FF} & \textbf{Freq. (MHz)} \\
\midrule
$4\times4$ & \Name & 3,505 & 4,291 & 406 \\
 & RTL & 2,309 & 3,378 & 463 \\
$8\times8$ & \Name & 14,863 & 18,453 & 381 \\
 & RTL & 9,694 & 15,332 & 336 \\
$16\times16$ & \Name & 54,439 & 72,585 & 329 \\
 & RTL & 57,499 & 91,251 & 311 \\
\bottomrule
\end{tabular}
\par\endgroup

\end{minipage}
\caption{FEATHER latency for (a) $128^3$ GEMM and (b) $3\times3$ convolution at spatial size $16\times16$ with 64 input/output channels. (c) Resident-weight compute-port resources with LUT-based multipliers.}
\label{fig:multikernel}
\Description{GEMM and convolution compare streamed SPMW with corrected RTL using the original PE-wise loader and improved row-wise loader. The resource panel compares resident-weight SPMW with LUT-based multiplication and PE-loader RTL at 4 by 4, 8 by 8, and 16 by 16. SPMW LUT/FF counts are 3505/4291, 14863/18453, and 54439/72585, versus RTL counts of 2309/3378, 9694/15332, and 57499/91251. Their routed frequencies are 406/463, 381/336, and 329/311 MHz, respectively. Both use zero DSPs, BRAM, and URAM, so these columns are omitted.}
\vspace{-10pt}
\end{figure}

We measure gate/up projection slices from LLaMA~3.2~1B~\cite{meta2024llama32} and a DeepSeek-V4-Pro expert~\cite{deepseek2026v4config}. These cover dense and expert FFN dimensions for the paired projections used in gated feed-forward layers, a structure shared by the LLaMA family~\cite{touvron2023llama2} and the dense and MoE Qwen3 models~\cite{qwen2025qwen3}. Each slice uses 64 tokens and 64 columns per projection, giving $(M,K,N)=(64,2048,128)$ and $(64,7168,128)$. We use int8 inputs and int32 accumulation, a common integer configuration supported by all three engines, to compare their data delivery and control under the same arithmetic. All engines use identical random inputs and shift-and-clip requantization.

Gemmini uses its weight-stationary, matmul-only configuration. We replace floating-point scaling with integer shift for the common epilogue and reorganize accumulator banks into 64-bit columns to enable FPGA RAM inference. VTA uses its Chisel \texttt{Core} with batch size one and shipped storage depths. Both baselines execute their native instruction sequences.

\begin{figure}[t]
\centering
\includegraphics[width=\columnwidth]{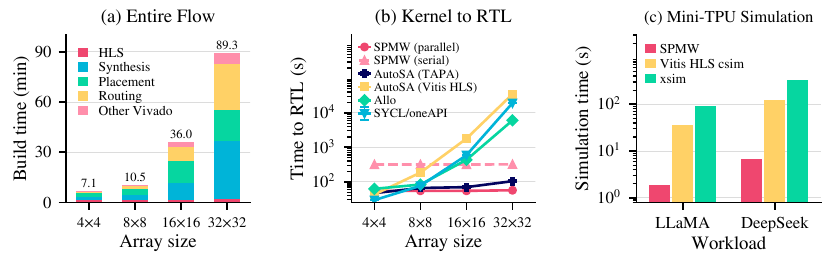}
\caption{Compilation and simulation time. (a) GEMM systolic array build time. (b) Compilation time from kernel to RTL. (c) LLaMA and DeepSeek simulation on a $16\times16$ mini-TPU.}
\label{fig:synth}
\Description{Three side-by-side plots. Stacked bars show HLS, synthesis, placement, routing, and other Vivado time. The second plot compares kernel-to-RTL compilation, including frontend, HLS, and RTL assembly. SPMW parallel times are 54.4, 54.5, 54.5, and 56.6 seconds; measured serial times are 315.9, 316.0, 316.0, and 319.2 seconds. AutoSA (TAPA) takes 46.8, 65.9, 69.8, and 104.0 seconds. Eight-worker SPMW and TAPA values are medians of three complete runs; serial SPMW values are single runs. AutoSA (Vitis HLS) and Allo add newly measured frontend times to their earlier HLS results. New curves omit error bars. SYCL includes the kernel frontend but excludes host compilation and final packaging, retaining its range bars. The third plot shows hot simulation times for LLaMA and DeepSeek gate/up slices on a 16-by-16 mini-TPU. SPMW uses non-fiber OpenMP execution with FIFO depth floor 64. Median times are 1.934 and 6.94 seconds for SPMW, 35.5735 and 124.5815 seconds for Vitis HLS csim, and 92.3 and 329.85 seconds for xsim. Simulation bars have no error bars.}
\vspace{-18pt}
\end{figure}

\figref[a,b]{fig:mini-tpu} compares whole-engine cycle latency and effective throughput for $n\times n$ arrays with $n=4,8,16$. Effective throughput is $2MKNf/C$, counting two operations per MAC, where $C$ covers instruction delivery through result writes and $f$ is the best reported routed frequency capped at 300~MHz. At $16\times16$, \Name's LLaMA slice uses 14.6\% and 11.6\% fewer cycles than Gemmini and VTA, respectively. \Name stays within 1.1\% of the arithmetic cycle bound $MKN/n^2$ by reusing weight blocks across input rows and overlapping operand fetching through fabric bindings. Its throughput is $2.01$--$2.32\times$ Gemmini's and $1.39$--$1.51\times$ VTA's, primarily because it reaches 300~MHz. \Name's critical path stays within a registered multiply-add cell and partial-sum link, whereas the baselines are limited by memory-access control and arithmetic. Gemmini's weight-tile loading and VTA's separate epilogue passes account for the cycle differences.

\figref[c]{fig:mini-tpu} reports resource usage for all three engines. At $16\times16$, \Name uses 46\% fewer LUTs than Gemmini. Its narrower GEMM instruction and blocked access pattern reduce control and storage costs. Against VTA, \Name uses similar LUTs and fewer RAM blocks but more FFs because it retains weights and array state in registers.

\reviewissue{R13}{\figref{fig:mini-tpu} and complete-engine resource rows use 98c8486 (3 October). Each plotted value uses whole-engine RTL cycles for a 64-token, 128-output gate/up slice with full K=2048 or 7168. No full-pair extrapolation. All 18 result records pass. SPMW/VTA consolidated cycles match individual xsim logs. Gemmini commits consolidated result lines and testbenches, not individual simulator logs. All selected P\&R values match raw timing and utilization reports. SPMW uses the 3.0 ns routes, Gemmini the 3.333/3.333/6.0 ns routes, and VTA the 4.6/4.6/4.3 ns routes. Each design selects its shortest period over the reported targets, then caps the frequency used for effective throughput at 300 MHz. Cycles are launch/first-instruction through completion after writes, with ideal command issue and 64-bit memory service. VTA read channels are served independently above its arbiter. Outputs are random-int8 projection slices, not native model-precision inference or accuracy results. Runtime reuse is checked using one hardware build per size and the source's 78-file identity records. Earlier fixed-function projection and SwiGLU rows remain archived. Normalization rows remain at 23f47a3, with E8 raw reports at 65bcbb7 for the file and Gemmini variants. Reload resources rely on the CSV. Retain the pending check to restore the softmax exponential in that Gemmini configuration. Hybrid E3 board results and source-line counts describe the earlier instruction engine.}

\textbf{FEATHER.}
FEATHER~\cite{tong2024feather} combines the NEST compute array with BIRRD, a programmable reduction and reordering network. The NEST array has height $AH$ and width $AW$. \figref{fig:feather-topology} shows BIRRD at $AW=8$, with six stages of four switches. The topology reverses selected lane-index bits between stages and connects paired-PE column sums to the first stage. \Name decouples weight delivery from these connections, using parallel weight streams with the same reduction network.

We evaluate $128^3$ GEMM and $3\times3$ convolution at spatial size $16\times16$ with 64 input/output channels on $n\times n$ arrays ($n=4,8,16$). The released RTL~\cite{feather_github} has a weight-buffer controller error, which we correct in both baseline variants. In \figref{fig:multikernel}, \emph{RTL} loads one PE weight per cycle, taking $n^3$ cycles despite its $n$-byte port. Our row-wise loader in \emph{RTL~(fixed)} uses all $n$ bytes but still serializes $n$ rows with $n$ weight slots per PE, taking $n^2$ cycles. Private weight files also require repeated delivery of weights shared across columns. \Name separates this delivery from PE behavior and reduction connectivity. Grid-derived boundary bundles expose each pair's weight port, which we bind to an independent stream to fill all pairs' $2n$ slots concurrently in $2n$ cycles. Matching this parallelism requires redesigning the RTL distribution interconnect and control beyond the row-loader fix. This $n/2$ loading advantage yields $2.00$--$7.84\times$ GEMM speedups over RTL~(fixed), while all three designs compute at an $n$-cycle interval with resident weights. Convolution pads nine taps to 12/16/16 and similarly achieves $2.00$--$7.98\times$ speedup.

\figref[c]{fig:multikernel} compares resident-weight compute ports with LUT-based multiplication and no DSPs, BRAM, or URAM. At $16\times16$, \Name uses 5.3\% fewer LUTs and 20.5\% fewer FFs, but $4\times4$ and $8\times8$ require 52--53\% more LUTs and 20--27\% more FFs. \Name's HLS control and FIFOs add overhead at small sizes, while RTL's register-based weight files and selection logic grow with $n$.
\reviewissue{R5}{The row-wise loader passes bit-exact arithmetic, weight-file checks, and full host GEMM reduction in feed mode. All six row-loader convolution rows at 1ef4599 pass tile checks, and all three feed-mode rows pass complete host convolution checks. Raw completion times are 3145745/2097179/1048621 cycles in feed mode and 786461/262227/65821 with resident weights. Resident tests reuse tile 0 weights. Its new P\&R rows at 23eedb2 have LUT 2291/9690/58465 and WNS +1.074/+0.386/+0.115 ns. \figref[c]{fig:multikernel} now uses the resident-weight SPMW fabric-multiply records and corrected PE-loader RTL at c20b2ee8, checked against all six routed utilization and timing reports. Both use zero DSPs, BRAM, and URAM. SPMW LUT/FF counts at sizes 4/8/16 are 3505/4291, 14863/18453, and 54439/72585, with WNS +0.869/+0.707/+0.291 ns. The resource cohort is distinct from the streamed latency cohort and excludes external operand supply. RTL includes its controller and narrow-port weight-delivery logic, whereas SPMW exposes parallel compute ports. This panel is not an equal-bandwidth full-system area comparison. The earlier DSP-inferred runs have WNS -0.699 ns for streamed SPMW 16x16 and +0.338 ns for its resident port, and are not the new LUT-multiplier resource rows.}

\subsection{Compilation and Simulation Time}
\label{sub:synth}
We measure compilation time for the systolic array experiment in \S~\ref{sub:compute}. \figref[a]{fig:synth} breaks down HLS through routing for one memory-connected build per size, using eight HLS workers and excluding frontend time. At $32\times32$, HLS takes 110.6~s, only 2.1\% of the 89.3-minute total. Preserving regularity lets the compiler reuse specialized PE kernels, keeping HLS small, while synthesis and P\&R still process the instantiated hardware.

\figref[b]{fig:synth} measures time from the source kernel to complete RTL, including the frontend, HLS, and RTL assembly but excluding IP packaging and P\&R. We additionally measure the TAPA~\cite{guo2023tapa} backend of AutoSA, which also uses parallel compilation for its tasks. We use eight workers for \Name and AutoSA (TAPA), and measure serial \Name with one worker. AutoSA (Vitis HLS) and Allo combine separately measured frontend and single-process HLS times, excluding HLS tool startup. We also add SYCL/oneAPI~\cite{intel_oneapi_pipes} for comparison with Altera's FPGA flow, using DPC++ 2025.0 for Agilex~7 at 300~MHz and timing the kernel frontend through RTL generation. As we do not have an Agilex~7 device, this experiment compares compilation time only.

At $32\times32$, \Name takes 56.6~s with eight workers, giving $619\times$, $107\times$, and $340\times$ speedups over AutoSA (Vitis HLS), Allo, and SYCL. Its measured 319.2~s serial run gives $110\times$, $19\times$, and $60\times$ speedups. AutoSA (Vitis HLS) and Allo process the full array in one HLS project. AutoSA (TAPA) reuses distinct synthesized tasks, but its RTL assembly grows to 39.2~s at $32\times32$, while \Name generates fabric RTL during frontend elaboration. \Name is still $1.84\times$ faster with eight workers. Parallel \Name compilation stays below 57~s, although AutoSA and SYCL are faster at $4\times4$.

\reviewissue{R4}{Panel (a) retains the memory-connected E1 builds, with fifteen kernel configurations and one P\&R run per size. Other Vivado time is the total minus synthesis, placement, and routing. Panel (b) uses 31673105 and total\_s for SPMW kernel and AutoSA-TAPA source-to-RTL walls, checked against runs.ndjson and results.csv. Eight-worker values are medians of three complete runs, not sums of stage medians. SPMW serial is a measured one-worker run. SPMW includes memory loaders and drain. Frontend, HLS, and RTL assembly are included; packaging and P\&R are excluded. AutoSA-Vitis and Allo add the new frontend measurements in frontends.ndjson to the 23eedb2 single-process csynth measurements, preserving the best-latency AutoSA interfaces. The frontend outputs match the E1 HLS input files byte for byte. These older HLS timers omit tool startup and project setup. At size 32, SPMW takes 56.6 s parallel and 319.2 s serial; TAPA takes 104.0 s and 425.1 s. AutoSA-Vitis and Allo totals are 35025.02 s and 6055.32 s. Earlier HLS-only curves remain archived. SYCL uses 2366056 (16 October), checked against all raw stage logs and NDJSON kernel summaries. Kernel-only times are 29.3, 76.1, 589.3, and 19222.7 s, taking HLS completion timestamps minus host compilation profiled once per size in separate runs. Three runs at sizes 4, 8, 16 and one at 32. All complete HLS successfully. Emulation passes through size 16, but size 32 times out at 1200 s. No FPGA execution, routed resources, or latency claim is made for SYCL. Simulation uses the October 23 records at 6268c1d4. The plotted workloads are E9's ptpu-llama and ptpu-dsv4 on 16x16 arrays, with K=2048 and K=7168, respectively. Both are 64-token int8 gate/up slices with 64 columns per projection. SPMW uses the two hot rerun\_s measurements at each size with FIFO depth floor 64 and non-fiber OpenMP threads. This simulation-only capacity setting reduces blocking and does not change the RTL configuration. Fiber runs are excluded. Vitis uses all four hot\_s values from two default csim\_design builds, with no call reordering. Vitis and xsim retain the original size-4/16 records and size-8 supplement. Xsim uses the separately timed xsim command. The size-4/16 records are checked against raw cosimulation logs, which check 8192 outputs and report 3670030 and 229426 RTL cycles. The size-8 supplement contains stage summaries, successful return codes, pass flags and 917530 cycles, but does not commit its raw cosimulation logs. SPMW and C simulation check functionality, without RTL cycle fidelity. Vitis timers include process startup and operand-file input, unlike SPMW's in-process invocation. The machine is shared. All run ranges remain archived although panel (c) now shows only medians. Measurements at the other array sizes, original FIFO depths, and fiber settings remain archived.}

\figref[c]{fig:synth} compares \Name's OpenMP simulator, which executes grid instances concurrently, with Vitis HLS csim and xsim on the $16\times16$ mini-TPU LLaMA and DeepSeek setup in \S~\ref{sub:programmable}. We time execution after compilation. The csim runs use the default debug executable. The xsim timer excludes HLS, testbench generation, and RTL compilation and elaboration. For LLaMA and DeepSeek, \Name takes 1.93~s and 6.94~s, giving $18.4\times$ and $18.0\times$ speedups over Vitis HLS csim and $47.7\times$ and $47.5\times$ over cycle-accurate xsim. This faster functional feedback lets designers identify computation and communication errors before running lengthy RTL simulations, shortening each correction and validation iteration.

\begin{table}[t]
\captionsetup{font=footnotesize,justification=raggedright,singlelinecheck=false}
\begin{minipage}[t]{0.425\linewidth}
\vspace{0pt}
\centering
\caption{Latency (cycles) for complete~$P[S,S]V[S,8]$~products.}
\label{tab:attention}
\footnotesize
\setlength{\tabcolsep}{3pt}
\begin{tabular}{@{}rrrr@{}}
\toprule
$\boldsymbol{S}$ & \textbf{Original} & \textbf{Grouped} & \textbf{Speedup} \\
\midrule
128 & 1,222 & 825 & $1.48\times$ \\
256 & 4,302 & 2,294 & $1.88\times$ \\
512 & 16,606 & 8,446 & $1.97\times$ \\
1,024 & 65,790 & 33,038 & $1.99\times$ \\
\bottomrule
\end{tabular}
\Description{Both mappings compute P of shape S by S times V of shape S by 8, using 256 PEs. For S of 128, 256, 512, and 1024, original measured latencies are 1222, 4302, 16606, and 65790 cycles; grouped measured latencies are 825, 2294, 8446, and 33038 cycles. Speedups are 1.48, 1.88, 1.97, and 1.99 times. Timings include all key tiles and feedback accumulation in one launch.}

\end{minipage}\hfill
\begin{minipage}[t]{0.555\linewidth}
\vspace{0pt}
\centering
\caption{Design-only source lines. Ratios use HLS/\Name, except FEATHER (RTL/\Name).}
\label{tab:loc}
\footnotesize
\setlength{\tabcolsep}{3pt}
\begin{tabular}{@{}lrrrr@{}}
\toprule
\textbf{Design} & \textbf{HLS} & \textbf{RTL} & \textbf{\Name} & \textbf{Ratio} \\
\midrule
Systolic array & 94 & -- & 24 & $3.92\times$ \\
FFT & 299 & 5,284 & 106 & $2.82\times$ \\
Mini-TPU & 673 & 3,226 & 154 & $4.37\times$ \\
FEATHER & -- & 1,802 & 97 & $18.58\times$ \\
\bottomrule
\end{tabular}
\Description{Source counts for four designs. FFT RTL counts the generated Spiral core. The mini-TPU row includes VTA HLS and Chisel Core under RTL. Ratios use HLS or, for FEATHER, RTL as the baseline.}

\end{minipage}
\end{table}

\subsection{Architectural Exploration}
\label{sub:reconfig}
We implement the architectural exploration in \S~\ref{sub:attnpv} and measure the latency benefit of reconnecting unused columns. Table~\ref{tab:attention} compares complete $P[S,S]V[S,d_h]$ products for $d_h=8$ and $S=128$--1,024 on two $16\times16$ arrays with 256 PEs each. Inputs are int8 and results int32. RTL timing spans reset release through the final output, including loading $V$ and accumulating all key tiles.

The original mapping holds 16 keys per tile in eight columns, with zeros in the other eight. Grouping joins the two column groups into eight reduction chains of depth 32, uses all 256 PEs, and halves the tile count from $S/16$ to $S/32$. Both designs preload all tiles' $V$ entries into each PE. After each tile, feedback FIFOs retain partial sums for all $S$ queries and return them to the chain inputs for the next tile. The final tile emits the complete product. Both designs drain pending results through flushable HLS pipelines during feedback stalls.

\looseness=-1
Grouping provides $1.48$--$1.99\times$ speedup, limited by pipeline fill and feedback at $S=128$. At $S\geq256$, both sustain one query per cycle per tile, so halving tiles approaches $2\times$. Adapting the RTL requires extensive changes to inter-group wiring and input delivery, while \Name only needs several lines of changes to fabric connections and boundary bindings without editing the MAC program.

\reviewissue{R6}{The complete-head experiment is pinned to implementation commit f6ca5bf5, with 31 source and result files hashed in attention\_head\_f6ca5bf5. Both designs compute the complete P[S,S] V[S,8] product with int8 operands and int32 results in one launch, using 256 MAC PEs. At S=64/128/256/512/1024, original cycle counts are 481/1222/4302/16606/65790 and grouped counts are 429/825/2294/8446/33038. All ten RTL runs pass against NumPy, with one seed per case. Table 2 reports the four sequence lengths from 128 through 1024. Original checks also include its padded zero output lanes. Both use flushable pipelines, per-PE V files loaded before the first query row, and feedback FIFOs of depth S to accumulate key tiles. The default pipeline style deadlocks the grouped S=64 case, which is archived but excluded from the comparison. The timings include initial loading, pipeline fill/drain, and feedback stalls from reset release through all outputs. No place-and-route frequency or area is established for these arrays. The former rectangular tile sweep and its estimated original totals remain archived and are no longer used in Table 2.}

\subsection{Productivity}
We assess productivity through source-size comparisons and a controlled coding-agent study.

\subsubsection{Code Size}
\label{sub:check}
Table~\ref{tab:loc} counts design source for systolic GEMM, FFT, mini-TPU, and FEATHER, excluding configuration, tests, software references, comments, and docstrings through language-aware filtering. FFT RTL is Spiral's 256-point, two-samples/cycle design~\cite{milder2012spiral}. Mini-TPU compares the \Name stage engine with VTA HLS and Chisel, counting Chisel as RTL~\cite{moreau2018vta}.

For systolic GEMM and FFT, \Name uses up to $3.92\times$ fewer source lines than HLS. A shared PE program replaces repeated arithmetic definitions, while grid and connection rules replace per-instance wiring. Boundary bindings also express operand delivery without spelling out each buffer and channel. FEATHER's reduction network needs 97 \Name lines versus 1,802 RTL lines using one switch program and connection rule.
\reviewissue{R7}{The author requested the original folded FFT count of 106 from the archived September 6 report, replacing the later 507-line lanes design. Spiral adds 5284 non-comment RTL lines from the 72 module definitions reachable from dft\_top in its generated N256 W2 core. The uninstantiated testbench and three unused helper modules are excluded. The source hash and in-place count are archived in spiral\_loc\_20261011; the generated RTL is not redistributed. HP-FFT UF1 remains 299 lines and carries two samples per cycle; the folded SPMW design carries one. These source-size ratios are not matched-throughput measurements. FEATHER RTL is freshly counted from the nine modules in the measured e4\_feather\_fixed/RTL directory: 1802 design lines, excluding eight timescale directives. The row-loader variant has 1810 lines. Source hashes and module selections are archived. RTL includes its controller, SRAM, and output bus; arithmetic and delivery differences remain described in Evaluation. The mini-TPU and other SPMW counts remain pinned at 23eedb2, with the full stage engine at 154 lines. VTA HLS is pinned at d4a15f6: 523 C++ design lines plus 150 Tcl hardware lines, excluding its simulation-only dispatcher, configuration/type aliases, test/host code, and generic board/build scaffolding. The Chisel Core count is 3226 design lines from the same d4a15f6 source, including default 64-bit memory-path dependencies and interfaces, with unused alternative module definitions, shells, tests, and configuration constants excluded. Source hashes and class selections are archived in \texttt{loc\_vta\_chisel\_20261011}. Ratios measure source length, not development time.}

\begin{figure}[t]
\centering
\begin{minipage}[c]{0.58\linewidth}
\includegraphics[width=\linewidth]{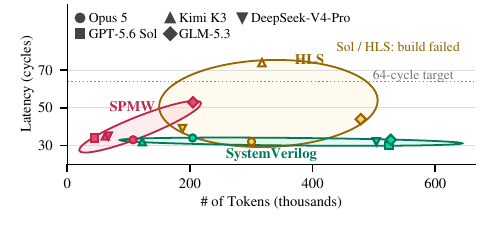}
\end{minipage}\hfill
\begin{minipage}[c]{0.405\linewidth}
\centering
\footnotesize
\setlength{\tabcolsep}{1.5pt}
\begin{tabular}{@{}lrrrr@{}}
\toprule
\textbf{Framework} & \makecell{\textbf{Meets}\\\textbf{latency}} & \makecell{\textbf{Tokens}\\\textbf{(K)}} & \textbf{Builds} & \makecell{\textbf{Time}\\\textbf{(min)}} \\
\midrule
\Name & 5/5 & 67.1 & 2 & 20.5 \\
C++ HLS & 3/5 & 300.4 & 8 & 37.1 \\
SystemVerilog & 5/5 & 503.8 & 4 & 26.6 \\
\bottomrule
\end{tabular}
\Description{Five trials per framework. Five SPMW, three HLS, and five SystemVerilog designs pass the twelve-product correctness tests and latency limit. Median costs include every trial, including build failure. The counts measure simulated correctness and latency.}

\end{minipage}
\caption{Agent study for an $8\times8$ array. Colors identify frameworks and symbols identify models. Filled markers meet the 64-cycle limit.}
\label{fig:agent-study}
\Description{The scatterplot groups SPMW, HLS and SystemVerilog measurements in colored ovals, with five model symbols. All measured designs are shown, including target misses, without a fabricated latency for the HLS build failure. The adjacent summary table reports 5/5, 3/5 and 5/5 passes, with median token costs 67.1K, 300.4K and 503.8K.}
\end{figure}

\subsubsection{Agentic Design}
\label{sub:agents}
We evaluate five coding models accessed through OpenRouter~\cite{openrouter2026}, Opus~5~\cite{anthropic2026opus5}, GPT-5.6 Sol~\cite{openai2026gpt56sol}, Kimi~K3~\cite{moonshot2026kimik3}, GLM-5.3~\cite{zai2026glm53}, and DeepSeek-V4-Pro~\cite{deepseek2026v4pro}, using \Name, C++ HLS, and SystemVerilog. Each model implements the same $8\times8$ output-stationary int8 systolic array with int32 accumulation, giving fifteen trials. The specification fixes 64 PEs, nearest-neighbor operand movement, and row-wise result draining. All implementations expose the same scalar stream ports. Every trial receives an interface stub, the same functional specification and testbench, and a language reference. The references have approximately equal token counts. Trials use temperature zero and nominal budgets of 1M billed tokens, eight builds, three routing attempts, and four hours. Each agent works in an isolated sandbox without access to the \Name source repository. It can edit designs, read supplied files, build, and route only through our provided APIs. The reported latency criterion requires exact outputs and first-product latency between 22 and 64 cycles, measured from the first input to the last output. The lower bound reflects nearest-neighbor propagation, while the upper bound rules out serial execution of all 512 MACs.

The left panel of \figref{fig:agent-study} compares total token cost with latency. All five \Name designs, three HLS designs, and all five SystemVerilog designs meet the latency criterion. \Name latencies are 33--53 cycles, compared with 30--34 for SystemVerilog and 32--74 for the four buildable HLS designs. GPT-5.6 Sol's HLS artifact fails to build, and Kimi's HLS design takes 74 cycles.

The right panel of \figref{fig:agent-study} reports median total tokens, builds, and elapsed time across all five trials per framework, including target misses and build failures. Elapsed time includes model calls and tool execution through trial stop, excluding subsequent evaluation. \Name reduces median token use by factors of $4.48\times$ and $7.51\times$ relative to HLS and SystemVerilog, respectively. It also reduces median elapsed time by factors of $1.81\times$ and $1.29\times$, respectively.

Post-trial routing meets the 300~MHz target for all fourteen buildable artifacts. Every \Name design uses exactly 64 DSPs. \Name needs a median of two builds. This suggests that decoupling PE behavior from replication and connectivity makes the design task easier for agents to express. SystemVerilog also yields five working designs, but explicit registers, wiring, and handshakes require more source construction. C++ HLS additionally requires coordinating stream types, array partitioning, and scheduling pragmas, which makes high-performance designs more challenging.
\draftnote{\textbf{Broader grading is complete.} The source study's full criterion also includes sustained interval, exactly 64 DSPs, a 25,000-LUT limit, and timing. It yields 2/5 SPMW, 2/5 HLS, and 4/5 SystemVerilog passes. Three SPMW and two HLS designs miss the interval limit. DeepSeek's SystemVerilog misses the DSP criterion. Preserve these outcomes in the evidence while the paper reports the author's requested latency criterion.}
\reviewissue{R8}{The 7c96b45 source review covers all fifteen designs and records conformance. DeepSeek's SystemVerilog uses logic multipliers and no DSPs, while every other buildable design uses 64 DSPs. The single trial per model/framework does not measure variance. The architectural explanation is an interpretation of observed outcomes, not a measurement of agents' internal understanding or prior language familiarity. This is manual inspection, not a proof from the cycle harness. The non-compiling HLS source remains a build failure. All fourteen buildable artifacts have routed reports. The earlier six-of-fourteen seeded structural-check result remains a separate experiment.}
\reviewissue{R9}{All fifteen latest rows match the twelve-product simulation log, and all fourteen routed rows match the stopped artifacts' utilisation and timing reports. The latency points, 5/5--3/5--5/5 latency counts, and all-trial cost medians are unchanged. The full-grading counts above additionally include interval and resource constraints and must not be substituted for the latency counts without changing their definition.}

\section{Discussion and Limitations}
\label{sec:discussion}

\textbf{Simulation and verification.}
\Name supports concurrent functional simulation, but performance exploration requires kernel synthesis and assembled-RTL simulation. Future work will reuse unit timing models across grid sizes and boundary schedules to provide fast performance simulation at the programming-model level. Moreover, structural checks and tests do not establish functional correctness or deadlock freedom for all inputs, which requires a formal verifier.

\textbf{Physical implementation.}
Logic synthesis and P\&R still process each full RTL configuration and dominate the largest array's build time (\S~\ref{sub:synth}). Our preliminary floorplanner improves frequency by mapping grid tiles and input FIFOs to neighboring FPGA clock regions and pipelining inter-die links, following the ideas of AutoBridge~\cite{guo2021autobridge} and RapidStream~\cite{guo2022rapidstream}. On a $32\times32$ matrix--vector engine on U280, it raises post-P\&R frequency by 32\%, from 346 to 457~MHz. Future work will jointly optimize partitioning and pipelined placement across accelerators and reuse placed-and-routed tiles to reduce rebuild time.

\textbf{Programming accelerators.}
The evaluated backend constructs FPGA hardware. Future work will map \Name programs onto AMD AI Engines~\cite{xilinxAIE,xdna_npu,mliraie,wang2026mlirair} or Cerebras~\cite{cerebras_wse}. We also plan to generate application compilers from \Name hardware specifications. Work-unit semantics and hardware contracts would guide instruction selection, tiling, and scheduling as hardware evolves.

\section{Related Work}
\label{sec:related}

\textbf{SPMD programming models.}
CUDA~\cite{cuda}, SYCL~\cite{khronos_sycl2020}, and Triton~\cite{tillet2019triton} organize communicating parallel programs over indexed execution instances. Cypress~\cite{rohan2025cypress} and Prism~\cite{bansal2026prism} improve GPU program composition through task/tensor mappings and typed execution perspectives, respectively. \Name makes replication, interconnect, and boundary supply explicit for hardware construction.

\textbf{Hardware DSLs and compilers.}
TAPA~\cite{chi2021tapa,guo2023tapa} connects C++ tasks through typed streams, supporting hierarchy, replication, kernel reuse, and parallel HLS compilation. Other flows optimize dataflow and data placement~\cite{yehpca2022scalehls,ye2024hida,suhail2025streamhls,ye2025streamtensor,xiang2022heteroflow,zhuang2025aries}, support dynamic streaming~\cite{ghosh2025ripple,sohn2026step}, or derive dataflow hardware from algorithms and control flow~\cite{zhang2026codo,li2026neuracompiler}. Hardware construction languages provide parameterized generators~\cite{bachrach2012chisel,spinalhdl,amaranth}, explicit control~\cite{rachit2021calyx}, resource-aware types~\cite{rachit2020dahlia}, and timing abstractions~\cite{durst2020aetherling,nigam2023filament,nigam2026lilac,yu2026anvil}. HLS tools such as XLS~\cite{google2026xls} and accelerator DSLs derive hardware from computations and mappings~\cite{vitis_hls_2023,koeplinger2018spatial,nitish2019t2s,lai2020susy,jason2018polysa,wang2021autosa,lai2019heterocl,fang2025dato,zhang2024pom}, while Allo~\cite{chen2024allo} decouples algorithms from hardware customization. \Name preserves how shared work units connect and receive data for structural checks and kernel specialization before RTL assembly. Topology changes recompute boundaries and port usage while reusing compatible HLS implementations.

\textbf{Coding agents for hardware.}
SWE-agent~\cite{yang2024sweagent} studies tool interfaces. VerilogEval~\cite{pinckney2024verilogeval} and Spec2RTL-Agent~\cite{yu2025spec2rtl} study hardware generation. CAKE~\cite{ye2026cake} co-designs a GPU schedule IR and agent workflow with compiler verification and localized diagnostics. For accelerator construction, \Name lets agents update all instances through shared declarations. Before HLS, the compiler detects inconsistent connections and missing boundary bindings.

\section{Conclusion}
\label{sec:conclusion}
\Name decouples work-unit behavior from grid connectivity and data delivery. We demonstrate its generality with fixed-function and domain-specific programmable accelerators. Shared contracts reduce repetitive edits for engineers and coding agents. Explicit connections enable early structural
checks, supporting faster exploration and composition as accelerators become more heterogeneous.

\ifdefined\SPMWArxiv\else
  \clearpage
  \input{sections/10-data-availability}
\fi

\bibliographystyle{ACM-Reference-Format}
\bibliography{spmw}

\end{document}